\documentclass[reprint,superscriptaddress,amsmath,amssymb,aps,prx]{revtex4-2}

\usepackage{graphicx}
\usepackage{dcolumn}
\usepackage{bm}
\usepackage{cases}
\usepackage{siunitx}

\usepackage[colorlinks=true,allcolors = blue]{hyperref}

\usepackage{newtxtext,newtxmath} 
\usepackage{microtype}

\usepackage{xcolor}
\newcommand{\change}[1]{{\color{black} #1}}

\newcommand{\kB}{k_{\mathrm{B}}}
\newcommand{\kT}{\kB T}

\newcommand{\myvec}[1]{{\mathbf{#1}}}
\newcommand{\Ivec}{\myvec{I}}
\newcommand{\Evec}{\myvec{E}}
\newcommand{\Uvec}{\myvec{U}}
\newcommand{\Bvec}{\myvec{B}}
\newcommand{\Hvec}{\myvec{H}}
\newcommand{\Jvec}{\myvec{J}}
\newcommand{\Dvec}{\myvec{D}}
\newcommand{\nvec}{\myvec{n}}

\newcommand{\lD}{\lambda_{\text{D}}}
\newcommand{\DDP}{D_{\text{DP}}}
\newcommand{\UDP}{\Uvec_{\text{DP}}}

\newcommand{\dl}{\mathrm{d}\myvec{l}}

\newcommand{\dS}{\mathrm{d}{S}}

\newcommand{\Da}{D_\mathrm{a}} 

\newcommand{\via}{\latin{via}}

\newcommand{\rion}{r_{\mathrm{ion}}}
\newcommand{\csalt}{c} 

\newcommand{\z}{\phantom{0}}
\newcommand{\zz}{\phantom{$-$}}

\newcommand{\msqpersec}{\mathrm{m}^2\,\mathrm{s}^{-1}}
\newcommand{\dunits}{$({}\times10^{-10}\,\msqpersec)$}

\newcommand{\Eq}[1]{Eq.~\eqref{#1}}
\newcommand{\Eqs}[1]{Eqs.~\eqref{#1}}
\newcommand{\Fig}[1]{Fig.~\ref{#1}}

\newcommand{\Refcite}[1]{Ref.~[\onlinecite{#1}]}
\newcommand{\Refscite}[1]{Refs.~[\onlinecite{#1}]}

\newcommand{\latin}[1]{{\itshape #1}}

\newcommand{\ie}{\latin{i.\,e.}}
\newcommand{\etal}{\latin{et al.}}

\newcommand{\caveat}{\latin{caveat}}

\newcommand{\numbr}[1]{({\it #1})}

\begin{document}


\title{Colloidal diffusiophoresis in crossed electrolyte gradients: experimental demonstration of an `action at a distance' effect predicted by the Nernst-Planck equations} 

\author{Ian Williams}
\affiliation{School of Mathematics and Physics, University of Surrey, Guildford GU2 7XH, United Kingdom}

\author{Patrick B. Warren}
\email{patrick.warren@stfc.ac.uk}
\affiliation{The Hartree Centre, STFC Daresbury Laboratory, Warrington WA4 4AD, United Kingdom}

\author{Richard P. Sear}
\email{r.sear@surrey.ac.uk}
\affiliation{School of Mathematics and Physics, University of Surrey, Guildford GU2 7XH, United Kingdom}

\author{Joseph L. Keddie}
\affiliation{School of Mathematics and Physics, University of Surrey, Guildford GU2 7XH, United Kingdom}

\date{\today}

\begin{abstract}
In an externally imposed electrolyte (salt) concentration gradient, charged colloids drift at speeds of order one micrometre per second.  This phenomenon is known as diffusiophoresis. In systems with multiple salts and `crossed' salt gradients, a nonlocal component of the electric field associated with a circulating (solenoidal) ion current can arise. This is in addition to the conventional local component that depends only on the local salt  gradients. Here we report experimental observations verifying the existence of this nonlocal contribution. To our knowledge this is the first observation of nonlocal diffusiophoresis. The current develops quasi-instantaneously on the time scale of salt diffusion. Therefore, in systems with multiple salts and crossed salt gradients, one can expect a nonlocal contribution to diffusiophoresis which is dependent on the geometry of the system as a whole and appears as a kind of instantaneous `action-at-a-distance' effect. The interpretation is aided by a magnetostatic analogy.   Our experiments are facilitated by a judicious  particle-dependent choice of salt (potassium acetate) for which the two local contributions to diffusiophoresis almost cancel, effectively eliminating conventional diffusiophoresis. This enables us to clearly identify the novel, nonlocal effect and may be useful in other contexts, for example in sorting particle mixtures. 
\end{abstract}


\maketitle

\section{\label{sec:intro}Introduction}
Diffusiophoresis (DP) is the directed transport of suspended particles driven by a gradient in solution concentration \cite{Marbach2019, Velegol2016}. Originally described by Derjaguin \cite{Derjaguin1947}, and developed by Prieve and Anderson \cite{Anderson1982, Prieve1984, Anderson1989}, DP is a consequence of the interaction between the particle surface and molecules in solution. In an externally-imposed concentration gradient, the particle-surface interfacial free energy depends on position and this induces motion towards where this free energy is lower \cite{Anderson1982}. This is known as chemiphoresis. If the solute in question is ionic, an additional electrophoretic contribution to particle motion arises in response to the electric field spontaneously established when anions and cations diffuse at different rates \cite{Prieve1984}.

Recent research has revealed the ubiquity of DP in processes ranging from biological transport \cite{Doan2020, Ramm2021} and pattern formation \cite{Alessio2023}, to dialysis \cite{Kar2014}, water purification \cite{Guha2015} and laundry detergency \cite{Shin2018}. Furthermore, researchers are harnessing DP to develop technologies in areas including self-stratifying coatings \cite{Schulz2020, Rees-Zimmerman2021}, enhanced oil recovery \cite{Shi2021}, and colloidal sorting and separations \cite{Shimokusu2020, Shin2020,Rasmussen2020,singh2020,chakra2023}.  Traditionally, researchers have experimentally and theoretically explored DP in linear, one-dimensional concentration gradients of electrolytes, charged nanoparticles, or neutral molecules \cite{Florea2014, Paustian2015, Shin2016, NeryAzevedo2017, Ault2017, Prieve2019, Wilson2020, Gupta2020, RamirezHinestrosa2021, Shah2022, Timmerhuis2022, ReesZimmerman2023}. Squires and coworkers have further demonstrated how colinear concentration gradients of multiple molecules can effect complex manipulation of suspended particles \cite{Shi2016, Banerjee2019}. However, gradients in any biological or technological context are unlikely to be so simple.

In a recent Letter \cite{Warren2020}, one of us (PBW) theoretically described DP in orthogonal concentration gradients of two different salts in two dimensions. For closed systems with one-dimensional gradients, there are electric fields but no electric currents. This is also true in two or three dimensions if there is only one salt. However, this work showed that two- or three-dimensional gradients of two or more salts will necessarily generate a nonlocal solenoidal current throughout the solution, with an associated electric field. This field is in addition to the local electric field present in  conventional one-dimensional DP. 

Local here means determined by local salt concentrations. Nonlocal means that as soon as orthogonal salt gradients are developed anywhere in the solution, the solenoidal current and hence the electric field appear everywhere in the solution. This electric field will move charged particles, and so this nonlocal DP behaves essentially as an almost instantaneous action at a distance. This is qualitatively different behaviour from local DP which only acts at a point in solution once the salt has diffused to that point. Here we present the first experimental evidence of this action-at-a-distance effect on colloidal particles.

The phenomenon may have applications in charge-sensitive colloidal sorting and separation, including electrode-free nanomaterial recovery and recycling \cite{Myakonkaya2010, Nazar2011, Zhang2019}. In addition to being much faster than conventional DP to start up, it acts to separate particles of different surface potentials as particles with different surface potentials follow different, diverging trajectories. Two-dimensional DP has further potential as an inexpensive and portable characterisation tool for concentrated suspensions, capable of extracting particle charge and isoelectric point from DP trajectories \cite{Rasmussen2020}. Precise control over colloidal trajectories by two-dimensional DP may also lead to advances in self-assembly by electrophoretic deposition \cite{Besra2007}.

\begin{figure}
\includegraphics[width=80mm]{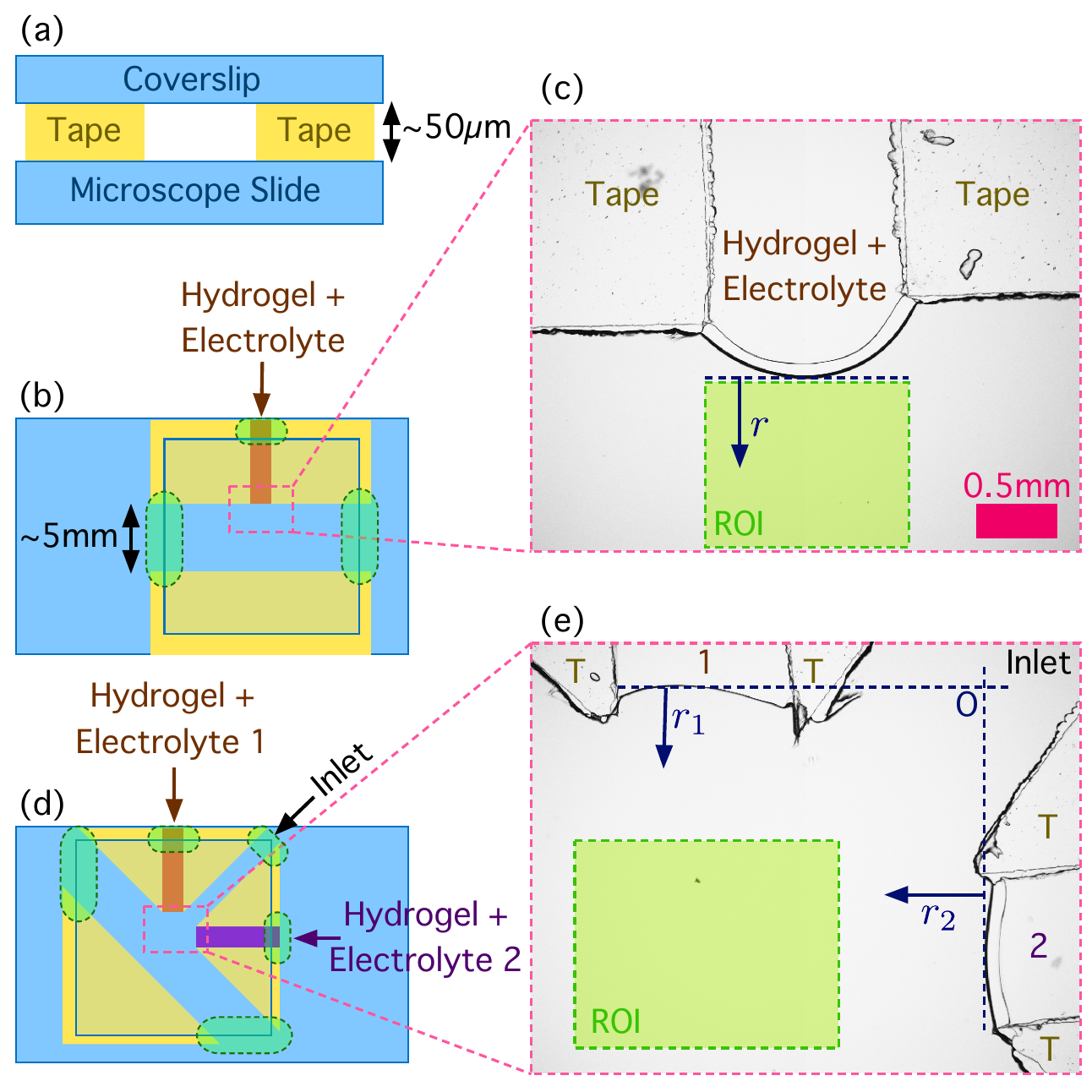}
\caption{\label{fig:Schematic} Microfluidic device schematics and micrographs. (a) Side view schematic showing microchannel construction using double-sided adhesive tape. (b) Top view T-shaped channel schematic for one-dimensional DP experiments. A poly(ethylene glycol) diacrylate (PEG-DA) hydrogel containing salt is formed in the source channel. \change{Green shaded ovals indicate the open channel ends which are sealed with glue after the experiment is initiated.} (c) Brightfield micrograph at $5 \times$ magnification showing the region close to the T-junction. Hydrogel and tape are labeled. \change{The navy blue dashed line indicates the hydrogel edge, defined as $r=0$, and the arrow indicates the co-ordinate, $r$, defined as the normal distance from the $r=0$ line. Green shaded region shows the region of interest considered for velocity profile calculations. (d) Top view schematic of the branched channel} for two-dimensional DP experiments. PEG-DA hydrogels containing salts are formed in perpendicularly oriented source channels. Sample inlet channel is located between the hydrogels and oriented at $45^\circ$. \change{Green shaded ovals indicate the open channel ends which are sealed with glue after the experiment is initiated.} (e) Brightfield micrograph at $5\times$ magnification showing the branch junction. Hydrogel sources are labeled 1 and 2. Tape walls are labeled T. Green shaded region indicates the region of interest for experiments performed at $10 \times$ magnification. \change{Navy blue dashed lines indicate the perpendicular hydrogel edges which form the axes of a Cartesian co-ordinate system, $(r_1,r_2)$, indicated by the arrows. The origin of this co-ordinate system is in the top right of the image and labeled O.} Scale bar in (c) also applies to (e).}
\end{figure}

Here we adapt the soluto-inertial beacons approach of Banerjee \etal\ \cite{Banerjee2016, Banerjee2019, Banerjee2019PRE, Banerjee2020} to realise electrolyte concentration gradients in microfluidic devices, as illustrated in \Fig{fig:Schematic}. We first characterise the DP of negatively charged polystyrene colloids in one-dimensional gradients of four salts using T-shaped microfluidic channels, \Fig{fig:Schematic}~(b,~c). We subsequently superpose orthogonal gradients of pairs of salts in branched devices, \Fig{fig:Schematic}~(d,~e), and provide the first experimental measurements of two-dimensional DP. We identify the current-induced contribution predicted by  \Refcite{Warren2020} and explore strategies to maximise its effect. 

\section{\label{sec:DPTheory}The Theory of Electrolyte Diffusiophoresis}
\subsection{\label{sec:1dDPTheory}Electrolyte Diffusiophoresis in 1d}
The Derjaguin-Prieve-Anderson (DPA) theory \cite{Prieve1984, Lee2022} of electrolyte DP predicts that a suspended particle placed into a gradient in salt concentration, $c$, will move with velocity
\begin{equation}
\label{eq:PrieveAnderson}
\UDP = \DDP \, \nabla \ln c\,.
\end{equation}
Assuming an ideal solution, a 1:1 electrolyte, zeta potential $\zeta$, and an infinitesimal double layer \cite{Prieve1984, Lee2022}, the DPA mobility is
\begin{equation}
\DDP =
\frac{\epsilon \kT}{\eta e} \left( \frac{4\kT}{e} \ln \cosh \frac{e\zeta}{4\kT}+\zeta \beta \right)\,.
\label{eq:mobility}
\end{equation}
Here $\epsilon$ and $\eta$ are the permittivity and viscosity of the solvent, $\kB$ is Boltzmann's constant, $T$ is absolute temperature and $e$ is the elementary charge. The parameter
\begin{equation}
\label{eq:beta}
\beta = \frac{D_+ - D_-}{D_+ + D_-}
\end{equation} 
encodes the difference in diffusion coefficients between the anion, $D_-$, and the cation, $D_+$. 

More general forms of \Eq{eq:mobility} can be formulated \cite{Prieve1984, Shin2016, Lee2022}. When concentration gradients lead to gradients in solution pH, it is necessary to account for the concentration dependence of $\zeta$ \cite{Lee2022, Shim2022}. Researchers have also further developed DP theory to account for multivalent electrolytes \cite{Wilson2020} or high ion concentrations \cite{Prieve2019, Gupta2020}. However, the majority of researchers have found the expression of \Eq{eq:mobility} and the assumption of constant $\zeta$ to be sufficient \cite{Kar2014, Paustian2015, NeryAzevedo2017, Banerjee2016, Boulogne2017, Battat2019, Shah2022}. 

According to the above, $\DDP$ is the sum of two contributions. The first term in \Eq{eq:mobility} is a chemiphoretic contribution arising as a consequence of the free energy of the particle surface depending on the local salt concentration. This term is always positive and independent of $\beta$, meaning that chemiphoresis is always directed up-gradient.  The second term in \Eq{eq:mobility} is a local electrophoretic contribution, which is a consequence of the electric field established when the anion and cation diffuse at different rates. This depends on the product $\zeta \beta$. Thus, depending on the sign of the particle surface potential and which ion diffuses more rapidly, the local electrophoretic contribution can be directed in either direction, up or down the gradient. 

Table~\ref{tab:Ds} shows $D_-$, $D_+$, and $\beta$ for the salts used in this research. These are tetrabutylammonium bromide (TBAB), sodium chloride (NaCl), potassium chloride (KCl), and potassium acetate (KOAc), chosen for the range of $\beta$ they collectively span. The contributions to $\DDP$ according to \Eq{eq:mobility} are given in Table~\ref{tab:DDP}, for $\zeta=-50\,\mathrm{mV}$, corresponding to the particles used in our experiments.

\begin{figure}
\includegraphics[width=80mm]{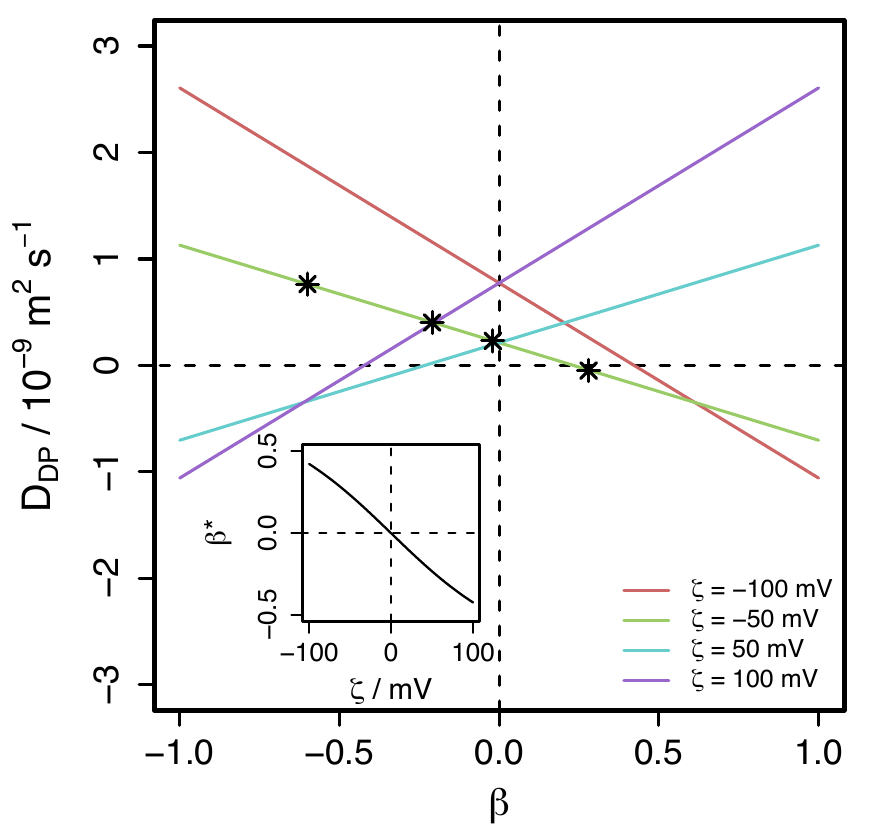}
\caption{\label{fig:mobility} DPA diffusiophoretic mobility, $D_{DP}$, as a function of the reduced difference in cation and anion diffusion constants, $\beta$ for particles with $\zeta$ potentials indicated in the legend. These are calculated according to \Eq{eq:mobility} for salts in water at 300\,K. The asterisks represent $\beta$ and $\DDP$ for the combination of the particles with $\zeta=-50\,\mathrm{mV}$ and salts employed in our experiments, (from left to right): TBAB, NaCl, KCl and KOAc. The inset shows the value of $\beta$ necessary for diffusiophoretic neutrality, $\beta^*$, defined as $\DDP(\beta^*)=0$, as a function of particle $\zeta$ potential.}
\end{figure}

Because the local electrophoretic term can have either sign, it can add to or partially cancel the chemiphoretic term, depending on the values of $\zeta$ and $\beta$. \Fig{fig:mobility} shows $\DDP$ as a function of $\beta$ for different values of particle $\zeta$ potential. When $\DDP > 0$ ($<0$), the net DP velocity is directed up (down) the concentration gradient. 

\begin{table}
\caption{\label{tab:Ds}Anion and cation diffusion coefficients at infinite dilution, $D_-$ and $D_+$, and the diffusivity difference parameter $\beta = (D_+ - D_-)/(D_+ + D_-)$. Data from \Refscite{Velegol2016} and \cite{CRC}.}
\vspace{1mm}
\begin{ruledtabular}
\begin{tabular}{rccc}
& $D_{-}$ & $D_{+}$ & \\
Electrolyte & \dunits & \dunits &  $\beta$ \\
\hline\\[-9pt]
TBAB & 20.8 & \z5.2 & $-$0.60 \\
NaCl & 20.3 &  13.3 & $-$0.21 \\
KCl  & 20.3 &  19.6 & $-$0.02 \\
KOAc & 10.9 &  19.6 & \zz0.28 \\
\end{tabular}
\end{ruledtabular}
\end{table}

\begin{table}
\caption{\label{tab:DDP}Theoretical $\DDP$ according to the DPA model, \Eq{eq:PrieveAnderson}, calculated with $\zeta = -50\,\mathrm{mV}$ and $\beta$ given in Table~\ref{tab:Ds}, assuming solvent viscosity $\eta = \SI{1}{\milli\pascal\second}$ and permittivity $\epsilon = 80\,\epsilon_0$. }
\vspace{1mm}
\begin{ruledtabular}
\begin{tabular}{rccc}
& \multicolumn{2}{c}{Contribution to $\DDP$ \dunits} & \\
 Electrolyte & Local Electrophoresis & Chemiphoresis & Total \\
\hline\\[-9pt]
TBAB & \zz5.5\z & 2.13 & \zz7.63  \\
NaCl & \zz1.9\z & 2.13 & \zz4.03  \\
KCl  & \zz0.17  & 2.13 & \zz2.3\z \\
KOAc & $-$2.6\z & 2.13 & $-$0.48  \\
\end{tabular}
\end{ruledtabular}
\end{table}

For a given $\zeta$, there is a value $\beta^*$ for which $\DDP=0$ (\Fig{fig:mobility}). This corresponds to the two terms inside the parentheses in \Eq{eq:mobility} perfectly cancelling, a condition we refer to as \emph{diffusiophoretic neutrality}.  If we linearise the $\ln\cosh$ term we obtain the approximate expression
\begin{equation}
\beta^*\approx -\frac{e\zeta}{8\kT}\,.
\end{equation}
The sign of $\beta^*$ depends on the sign of the charge on the particle, and the magnitude increases with its $\zeta$ potential.
The value for $\beta^*$ as a function of $\zeta$ obtained from \Eq{eq:mobility} is plotted in
the inset to \Fig{fig:mobility}. For example, KOAc has $\beta = 0.28$ and is approximately DP neutral when particles have $\zeta = -50\,\mathrm{mV}$. But, if particles have the opposite charge, $\zeta = 50\,\mathrm{mV}$, NaCl with $\beta = -0.21$ is a better choice for DP neutrality.

\subsection{\label{sec:2dDPTheory}Electrolyte Diffusiophoresis in 2d}
To the best of our knowledge, DP in two dimensions when perpendicular gradients of two salts are superposed has been described only in \Refcite{Warren2020}.  The theory predicts a DP velocity given by
\begin{equation}
\begin{split}
\UDP = &  \frac{\epsilon}{\eta} \Bigl( \frac{\kT}{e} \Bigr)^2\,\Bigl[4 \ln \cosh \left( \frac{e \zeta}{4\kT} \right) \nabla \ln \sum_i c_i  \\
  & \hspace{8em}{}
  + \frac{e \zeta}{\kT} \frac{\nabla g}{\sigma} + \frac{e \zeta}{\kT} 
  \frac{\Ivec}{\sigma} \Bigr]\,,
  \end{split}
\label{eq:2ddp}
\end{equation}
where the index $i$ labels each ion species. Here ${g = \sum_i z_i D_i c_i}$  is a weighted sum of ion densities ($z_i=\pm1$ is the valence) and ${\sigma = \sum_i z_i^2 D_i c_i}$ is the conductivity.

The second and third terms in the above are obtained by considering the ion current $\Ivec$, which in terms of the dimensionless electrostatic potential $\varphi$ and corresponding electric field $\Evec = -\nabla \varphi$, is given by
\begin{equation}
  \Ivec = -\nabla g + \sigma \Evec\,.\label{eq:Ivec}
\end{equation}
This result is derived by summing the set of Nernst-Planck equations (one per species)  which govern how the concentrations $c_i$ of the salt ions change with time \cite{Warren2020, newman2021}.  Inverting \Eq{eq:Ivec}, one can therefore associate the electric field with a local contribution from a gradient in $g$, and the current,
\begin{equation}
  \Evec =\frac{\nabla g}{\sigma} + \frac{\Ivec}{\sigma}\,.\label{eq:efield}
\end{equation}
Injecting this into the standard electrophoresis model gives rise to the second and third terms in \Eq{eq:2ddp} (the first term is just chemiphoresis, as in the one-dimensional case).  With this decomposition, the first and second terms of \Eq{eq:2ddp} are generalisations of the chemiphoretic and local electrophoretic terms of the standard DPA theory to the case of multiple salts. Both are local in the sense that they are functions {\em only} of the local gradients in the ion concentrations ($\nabla c_i$). The new (third) term cannot be written as a function only of local gradients and is therefore nonlocal.  Because this term is also fundamentally electrophoretic in origin, it scales with $\zeta$ the same way as the second term.

\subsubsection{Conditions for currents and hence action-at-a-distance diffusiophoresis, within the Nernst-Planck equations}
\label{subsub:action}
What are the conditions for a non-vanishing ion current?  First, imposing electroneutrality in the Nernst-Planck equations shows that the current must be \emph{solenoidal} \cite{Warren2020}, $\nabla\cdot\Ivec = 0$, which implies $\nabla\cdot(\sigma\nabla\varphi) + \nabla^2g = 0$ \cite{Warren2020, newman2021}. This is an inhomogeneous Poisson equation for the electrostatic potential, which replaces the true electrostatic Poisson equation in the problem (obviated by electroneutrality).  One can show \cite{Warren2020} that solutions $(\Ivec, \varphi)$ are uniquely specified up to an additive constant in $\varphi$ by the ion current through the boundaries (typically, in a closed system, $\Ivec\cdot\nvec=0$ where $\nvec$ is the surface normal).

Taking the `curl' of \Eq{eq:efield} and using the fact that $\nabla\times\Evec = 0$ (because $\Evec=-\nabla\varphi$) obtains
\begin{equation}
\nabla\times\left(\varrho\Ivec\right)=
\nabla g\times\nabla\varrho\,,
\label{eq:curlI}
\end{equation}
where $\varrho\equiv\sigma^{-1}$ is the resistivity.  This implies that if the gradients in $g$ and resistivity $\varrho$ (equivalently conductivity $\sigma$) are everywhere parallel (or antiparallel), then a solution exists with $\Ivec=0$ everywhere, and by virtue of the uniqueness theorem, this is the only solution. 

Thus $\Ivec\ne0$ requires `crossed' gradients $\nabla\sigma\times\nabla g\ne0$ \emph{somewhere} in the system.  However this does not imply that a current is restricted to such regions. It follows from \Eq{eq:curlI} that in regions where there are no gradients in $\varrho$ and $g$, the current is \emph{irrotational} \cite{Warren2020}; the current satisfies $\nabla\times\Ivec=0$. So in the region without crossed gradients, the conditions are $\nabla\times\Ivec=0$ and $\nabla\cdot\Ivec=0$. They can be met if $\Ivec$ is the gradient of some `current potential', $\Ivec=\nabla\omega$, which is harmonic ($\nabla^2\omega=0$) in the region in question.  

\begin{table}
\caption{Magnetostatic analogy for the ion current in the Nernst-Planck equations. Note that $\Bvec=\mu\Hvec$.  The boundary condition corresponds to a perfect conductor.  The notation is from Jackson's {\it Classical Electrodynamics}, 3rd ed.\ \cite{jackson1999}.\label{tab:mag}}
\vspace{1mm}
\begin{ruledtabular}
  \begin{tabular}{lll}
    Nernst-Planck & \multicolumn{2}{l}{magnetostatic analogy} \\
    \hline\\[-6pt]
    $\nabla\cdot\Ivec = 0$ & $\nabla\cdot\Bvec = 0$  & Gauss' law \\[3pt]
    $\nabla\times(\varrho\Ivec) = \nabla g\times\nabla\varrho$ &
    $\nabla\times\Hvec=\Jvec$ & Amp\`ere's law \\[3pt]
    $\Ivec\cdot\nvec = 0$ & $\Bvec\cdot\nvec=0$ & boundary condition\\
\end{tabular}
\end{ruledtabular}
\end{table}


That a non-vanishing, irrotational, solenoidal current is perfectly possible can also be seen by an unexpectedly close analogy (Table~\ref{tab:mag}) to magnetostatics \cite{jackson1999}.  This means that the ion currents around a region where there are crossed salt gradients look like the magnetic flux lines around a current-carrying conductor in the same place, and in two dimensions the ion current circulates around the crossed gradients like magnetic field lines encircling a current-carrying wire \footnote{There are some restrictions: \Eq{eq:efield} implies $\oint\!\varrho(\Ivec+\nabla g)\cdot\dl=0$.  Hence $\oint\!\Ivec\cdot\dl=0$ along a path where $\varrho$ is constant and $\oint\!\varrho\Ivec\cdot\dl=0$ along a path where $g$ is constant.  Since $\varrho>0$ this implies a circulating ion current must pass, possibly separately, through regions where $\varrho$ and $g$ are spatially varying \cite{Warren2020}. This result can also be derived by observing that $\nabla g\times\nabla\varrho = \nabla\times(g\nabla\varrho) = -\nabla\times(\varrho\nabla g)$ which implies \via\ Stokes' theorem $\int(\nabla g\times\nabla\varrho)\,\cdot\nvec\dS=\oint\! g\nabla\varrho\cdot\dl=-\oint\!\varrho\nabla g\cdot\dl$.  Hence, if there are net crossed gradients within an area $S$, there must be places on the perimeter where $\nabla\varrho\ne 0$ and where $\nabla g\ne 0$.}.  Note that just as magnetostatics demands $\nabla\cdot\Jvec=0$, the source term in \Eq{eq:curlI} is divergence-free, $\nabla\cdot(\nabla g\times\nabla\varrho)=0$\,; this follows from standard vector calculus identities \cite{jackson1999}.

In the present problem therefore, provided there are crossed gradients (in the sense that  $\nabla\sigma\times\nabla g\ne0$) somewhere in the system, one expects generically that an ion current will appear (quasi-instantaneously, see below) throughout the whole system, even in places where there are no crossed gradients.  According to \Eq{eq:2ddp} this ion current drives nonlocal DP, which should therefore in principle be observed quasi-instantaneously throughout the system.  A point to note, which will be important in the sequel, is that the effect should be largest where the conductivity is least, because this is where $\Evec$ is largest; this follows immediately from \Eq{eq:efield}.

The Nernst-Planck equations describe the evolution of the salt gradients, so we are always looking at the current distribution as a snapshot in time, as it were.  The magnetostatic analogy evolves as the salt gradients evolve, but it remains magneto-\emph{static}.  There is no equivalent to Maxwell's displacement current, the $\partial\Dvec/\partial t$ term in Maxwell's equations.

In any system with one-dimensional gradients all gradients are parallel and there are no currents, although the gradients may be coupled \cite{Gupta2019}.  It is also true however, that the current vanishes when there is only one salt present. This follows from electroneutrality. For example, for a 1:1 electrolyte, $\sigma=(D_1-D_2)c$ and $g=(D_1+D_2)c$, because electroneutrality forces $c_1=c_2=c$. Hence $\nabla\sigma\times\nabla g = 0$ everywhere. To summarise: in strictly one-dimensional systems, {\em or} when there is only one salt present, there is no current. However, if there are two or more salts present in a geometry that is {\em not} one-dimensional, `crossed' gradients are likely to arise, if not initially then almost certainly as the concentrations relax, making the accompanying ion currents and action-at-a-distance diffusiophoresis inevitable.

\subsubsection{Timescales in experiments in two-dimensional diffusiophoresis}
Before we discuss our experimental results, it is useful to consider the time scales relevant to our experimental system and elucidate precisely what is meant by `quasi-instantaneously' in the above discussion. The details are in Appendix~\ref{app:timescales}; here we summarise key conclusions. The longest time scales, and the only ones accessible in experiment, are that of diffusion of salt across the system. Our system is of order $L\approx\SI{1}{\milli\metre}$ across, so given the ambipolar salt diffusion coefficient $\Da\sim10^{-9}\,\msqpersec$, the time scale is $L^2/\Da\approx 10^3\,\SI{}{\second}\approx16\,\mathrm{minutes}$.  As we show in Appendix~\ref{app:timescales}, salt diffusion sets the time scale of both local diffusiophoresis terms, \ie\ including the local-diffusiophoresis contribution to the electric field. So all local DP propagates across our system at the speed of salt diffusion.

In \ref{subsub:action} we showed that the Nernst-Planck equations predict that nonlocal DP starts as soon as there are crossed salt gradients somewhere in the system. These salt gradients are established by diffusion, and so for salt sources $d$ apart, the gradients cross and nonlocal DP starts up after a time $d^2/\Da$. Our experimental setup is in \Fig{fig:Schematic}(e). The distance between the sources is about 2.5 times smaller than the distances from the sources to the farthest point from the sources imaged (bottom left of green shaded region). So nonlocal DP will reach the bottom-left point about five times faster than local DP.

Nonlocal DP starts, in effect, as soon as salt gradients cross because the nonlocal currents and electric fields are established and propagate very rapidly. It is this that makes the nonlocal DP look like instantaneous action-at-a-distance. The two relevant timescales are shown in \Fig{fig:timescales}, and are discussed by Bazant \etal\ \cite{bazant2004}, and in Appendix~\ref{app:timescales}. The timescales are for establishment of a space charge and charging of the electric double layers (EDLs). Both involve moving relatively tiny amounts of charge density, much less than that in the salt solution ($ce$). So they are much faster than salt diffusion and much faster than we can observe with microscopy. The space charge is needed to establish electric fields, and so the current. And the EDLs control the boundary conditions for these electric fields.

\begin{figure}
\includegraphics[width=80mm]{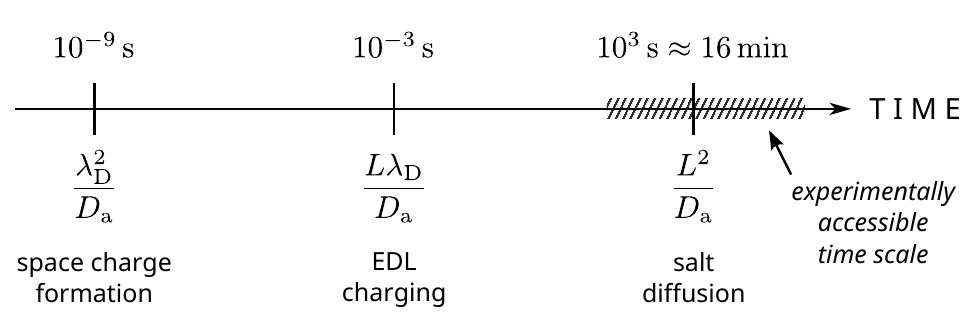}
\caption{Schematic of diffusion-limited time scales relevant to the system, from smallest (left) to largest (right). The shaded region is approximate range accessible in our experiments.  We assume the Debye length $\lD\sim10^{-9}\,\mathrm{m}$ ($100\,\mathrm{mM}$ salt), ambipolar diffusion coefficient for ions $\Da\sim10^{-9}\,\msqpersec$, and system size $L\sim10^{-3}\,\mathrm{m}$.  Details are given in Appendix~\ref{app:timescales}.}\label{fig:timescales}
\end{figure}

\begin{figure*}
\includegraphics[width=120mm]{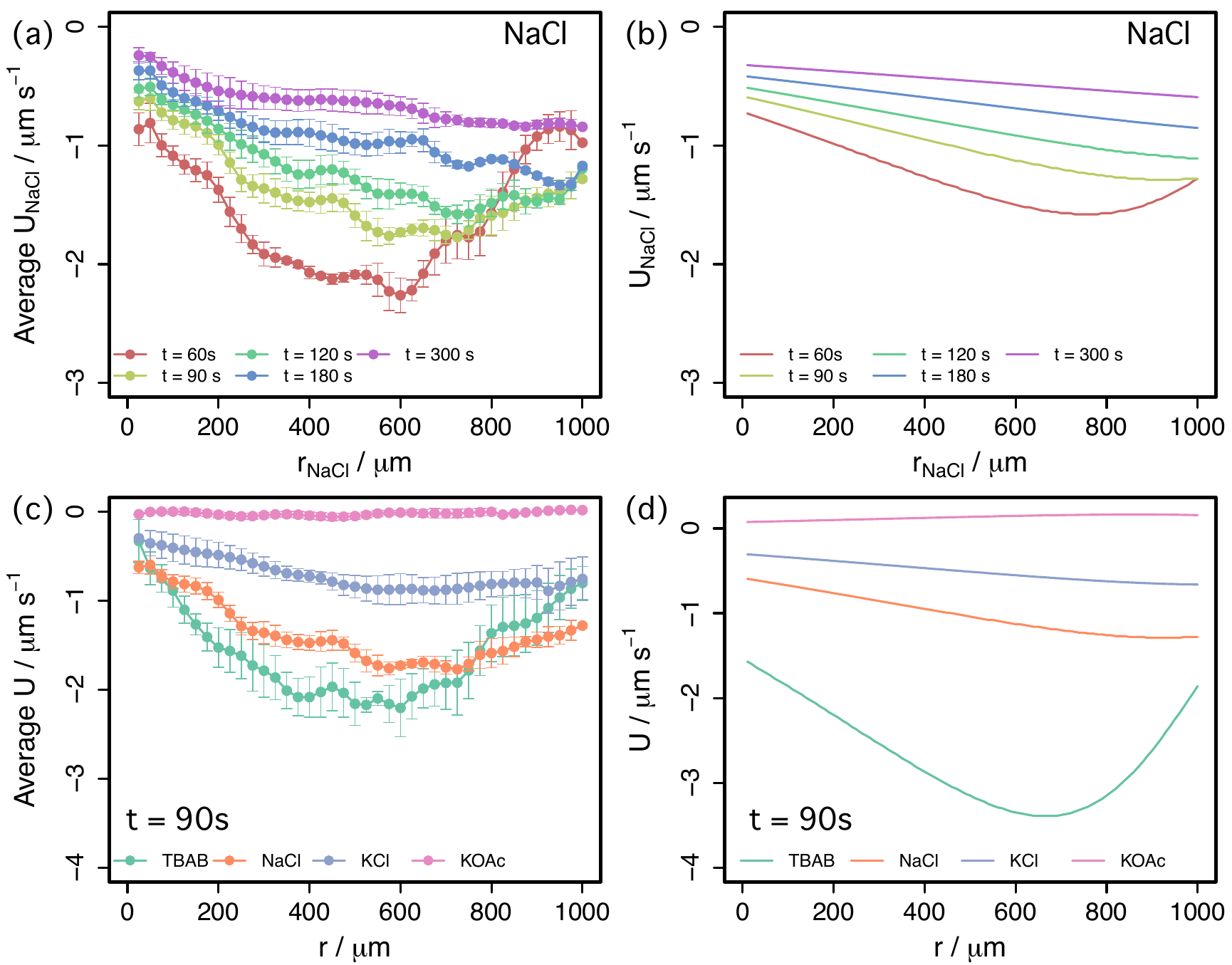}
\caption{\label{fig:1dDP} (a) Average DP velocity towards / away from a hydrogel source containing $130\,\mathrm{mM}$ NaCl as a function of distance from the source, measured in experiments. Colour indicates time elapsed between loading the the microfluidic device and data acquisition. Each data set is the average of three repeated experiments. Error bars indicate the standard error in the mean. (b) Theoretical DP velocity towards / away from an NaCl source as a function of distance from the source according to the DPA model, \Eq{eq:PrieveAnderson}. Colour indicates time elapsed since gradient initiation. (c) Average DP velocity towards / away from a hydrogel source as a function of distance from the source measured in experiment at $t = 90\,\mathrm{s}$ after loading. Sources contain $130\,\mathrm{mM}$ of different electrolytes, as indicated in the legend. Each data set is the average of three repeated experiments. Error bars indicate the standard error in the mean. (d) Theoretical DP velocity towards / away from electrolyte sources at $t = 90\,\mathrm{s}$ after gradient initiation as a function of distance from the source according to the DPA model, \Eq{eq:PrieveAnderson}. Colour represents electrolyte, as indicated in the legend.}
\end{figure*}

\section{\label{sec:results}Results and Discussion}
We seek to directly observe the nonlocal, current-induced electrophoretic contribution to DP predicted in \Refcite{Warren2020}. To this end, soluto-inertial beacons \cite{Banerjee2016} are used to create electrolyte gradients. Fluorescence microscopy and particle imaging velocimetry are employed to observe and quantify the motion of negatively charged colloids in these gradients. \change{Particle velocities are interpreted as arising exclusively as a result of diffusiophoresis, although in general we expect an additional contribution due to particle advection in the fluid flows generated by diffusioosmosis at the channel walls \cite{Gu2018,Williams2020}. This is discussed in greater detail in Section \ref{sec:conclusion}.}

Working in the microfluidic devices illustrated in \Fig{fig:Schematic}, we first measure DP in one-dimensional gradients of the four salts summarised in Tables~\ref{tab:Ds} and~\ref{tab:DDP}, and compare these experiments to the predictions of the DPA model. Subsequently, we superpose orthogonal gradients of pairs of these salts and measure DP in the two-dimensional concentration field. These experiments are compared to simulations implementing the model described in \Refcite{Warren2020}. Full methods are provided in Appendices \ref{sec:materials}, \ref{sec:microfluidics} and \ref{sec:microscopy}. 

\subsection{\label{sec:1dDPresults}Diffusiophoresis in 1d Salt Gradients}
One-dimensional experiments use T-shaped microfluidic devices with integrated hydrogel sources containing $130\,\mathrm{mM}$ of salt, as described in Appendix~\ref{sec:microfluidics} and shown in \Fig{fig:Schematic}~(b,~c). Particle velocity fields are estimated by PIV, as described in Appendix~\ref{sec:microscopy}, and the components of velocity towards / away from the salt source are spatially averaged to obtain velocity, $U$, as a function of \change{the normal} distance from \change{a line identified at the edge of the hydrogel, $r$, as indicated in Fig. \ref{fig:Schematic} (c). The region of interest is restricted to be directly in front of the hydrogel (green shaded region).} Negative velocities indicate motion up-gradient. The experiment is repeated three times with each salt and velocity profiles are averaged over the three repeats. 

The salts considered span a range of $\beta$. Tetrabutylammonium bromide (TBAB) and sodium chloride (NaCl) have $\beta < 0$ and are expected to drive DP in the same direction, with TBAB driving faster DP due to its larger absolute value of $\beta$. Potassium chloride (KCl) has $\beta \approx 0$ and therefore its DP is anticipated to be dominated by chemiphoresis. Potassium acetate (KOAc) has $\beta > 0$, meaning that its local electrophoretic component of DP is predicted to be oppositely directed to that due to TBAB or NaCl. To predict $U(r)$ for each salt using \Eq{eq:PrieveAnderson}, the time evolution of the salt concentration is estimated as described in Appendix~\ref{sec:app1d_diff}. 

Results are shown in \Fig{fig:1dDP}. The time evolution of the velocity profile measured in a NaCl gradient is shown in \Fig{fig:1dDP}~(a), and model predictions are in \Fig{fig:1dDP}~(b). Since our particles are negatively charged, DP speeds are always negative, \ie\ towards the source of the salt. Analogous plots for TBAB, KCl, and KOAc are in the Supplementary Information \cite{supplement}. 

The DP velocity profile changes with time because the salt concentration gradient is transient. Fast DP is measured at early times when the concentration gradient is steep but short-ranged. Initially, a peak in DP speed is measured at finite $r$. Especially at early times, far from the source, DP speed drops to zero as it takes time for salt to diffuse outwards and establish the local concentration gradient necessary to drive DP. The time for the salt to diffuse a distance $r$ is of the order $r^2/\Da$, which for a distance of $r\approx\SI{1000}{\micro\metre}$ is of the order ten minutes. We show in Appendix~\ref{app:timescales} that this local DP propagates at the speed of salt diffusion. At longer times the gradients weaken and so does DP. The trends in the experimental data are also seen in the model calculations in \Fig{fig:1dDP}~(b).

Experimentally measured DP velocity profiles for different salts at a single time ($t=90\,\mathrm{s}$) are compared in \Fig{fig:1dDP}~(c), with the corresponding theoretical predictions in \Fig{fig:1dDP}~(d). There is good agreement between the model and experiment. Salts with larger $\DDP$ drive faster DP, as expected. The chemiphoretic term is positive and identical for all monovalent electrolytes, meaning that chemiphoresis is predicted to always act up-gradient. Differences between salts are due to differences in their local electrophoretic terms, which depend on $\beta$. 

For TBAB and NaCl, the local electrophoretic contribution is positive, and acts in the same, up-gradient direction as chemiphoresis. $\DDP$ for TBAB is approximately twice as large for NaCl, and therefore, in identical gradients, TBAB is predicted to drive DP at approximately twice the speed of NaCl. The difference measured in experiment is not so large, but the measurements do show faster DP with TBAB. We speculate that \change{the main source of this} discrepancy is due to the unknown association and diffusion constants for the ions in the hydrogel. If TBAB associates with the hydrogel more strongly than NaCl, it will be emitted more slowly, meaning that the effective source concentration will be smaller for TBAB, and the concentration gradient will be more shallow. 

\change{An additional source of quantitative discrepancy between the experiments and modeling may be that the model is strictly 1-dimensional, while the experiments are not. The finite width and often slightly curved shape of the hydrogel surfaces, both mean that the iso-concentration contours in experiments will not be perfectly parallel to the line chosen as $r=0$.}

KCl has $\beta \approx 0$ as its anion and cation have very similar diffusion coefficients. Consequently, the local electrophoretic term for KCl is very small and KCl is predicted to drive DP primarily by chemiphoresis. The total $\DDP$ for KCl is still positive but smaller than for TBAB and NaCl. This is consistent with the experiments, which show up-gradient motion at slower speeds than both NaCl and TBAB.

\begin{figure*}
\includegraphics[width=130mm]{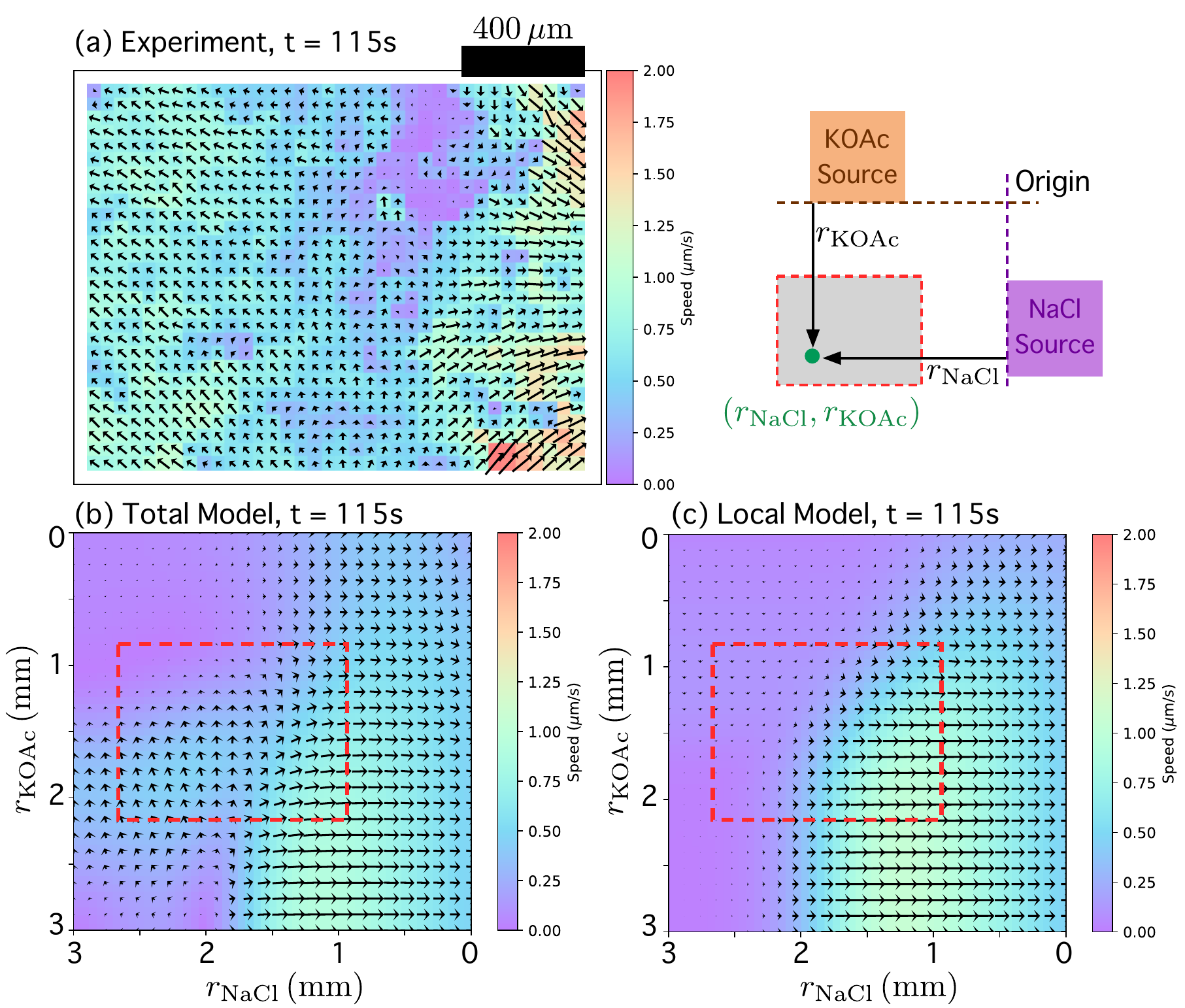}
\caption{\label{fig:2dArrows} (a) Experimentally measured velocity field in orthogonal concentration gradients of KOAc and NaCl at $t \approx 115\,\mathrm{s}$ after initiation. The KOAc source is located
  $\approx\SI{840}{\micro\metre}$ 
  from the top of the field of view and the NaCl source is located
  $\approx\SI{865}{\micro\metre}$ 
  from the right of the field of view, as indicated in the diagram on the right. The spacing between velocity vectors is
  \SI{44}{\micro\metre}.  
  (b,c) Simulated velocity fields at $t=115\,\mathrm{s}$ computed using (b) the full model, including nonlocal electrophoresis, and (c) the local model including only chemiphoresis and local electrophoresis. The red dashed rectangle corresponds to the location of the experimental field of view. \change{Top right illustrates the Cartesian co-ordinate system in which position is defined by the normal distances from lines drawn at the edges of the two sources and the origin is located between the sources.}}
\end{figure*}

\begin{figure*}
\includegraphics[width=175mm]{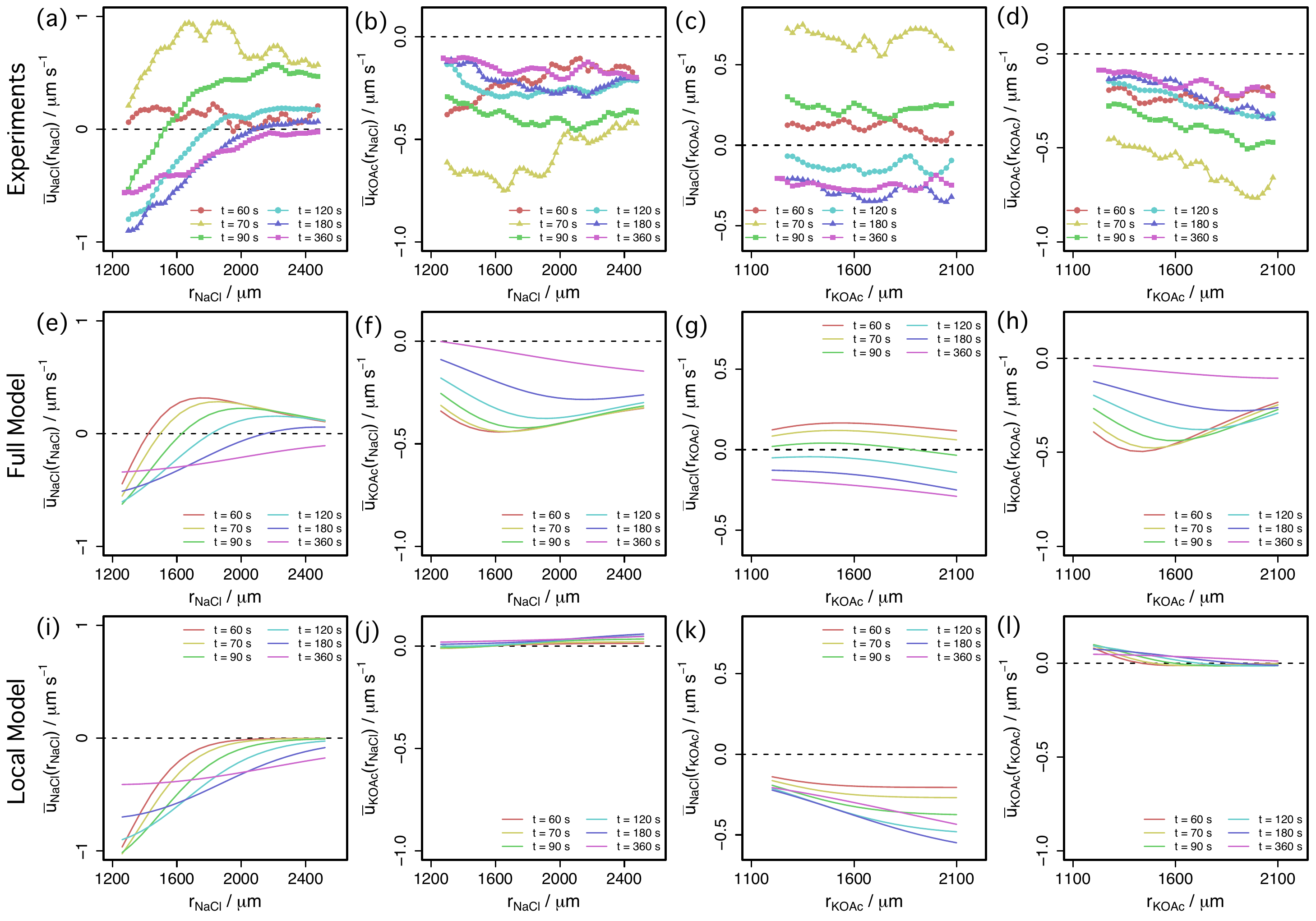}
\caption{\label{fig:KOAcNaClVProfs} KOAc superposed with NaCl. Projected velocity profiles in the region for which all three experiments overlap. Colour represents time after initiation, as indicated in legends. Horizontal dashed lines indicate $U=0$. (a--d) Experimental velocity profiles computed as an average over three repeated experiments. (e--h) Velocity profiles predicted by the full model, including the nonlocal contribution to DP. (i--l) Velocity profiles predicted by the local model:
  (a), (e), and (i) show $\bar{U}_\mathrm{NaCl}(r_\mathrm{NaCl})$\,;
  (b), (f), and (j) show $\bar{U}_\mathrm{KOAc}(r_\mathrm{NaCl})$\,;
  (c), (g), and (k) show $\bar{U}_\mathrm{NaCl}(r_\mathrm{KOAc})$\,;
  (d), (h), and (l) show $\bar{U}_\mathrm{KOAc}(r_\mathrm{KOAc})$.}
\end{figure*}

The local electrophoretic contribution for KOAc is negative, oppositely directed to chemiphoresis. The two terms have comparable magnitude and virtually cancel one another out, making KOAc an almost DP-neutral salt for our particles. DP is predicted to be an order of magnitude slower with KOAc than for the other salts, and indeed we observe negligible motion in a one-dimensional KOAc gradient. 

In summary, we have measured the DP motion of negatively charged particles under gradients of TBAB, NaCl, KCl, and KOAc. Experimental measurements are in good agreement with the predictions of \Eq{eq:PrieveAnderson}. We therefore consider the one-dimensional DP driven by these salts to be established and understood. If the objective is to drive fast DP in one dimension, then (for particles with $\zeta = -50\,\mathrm{mV}$) KOAc is almost the worst possible salt. However, this feature makes KOAc almost perfect for studying DP effects that only occur with superposed orthogonal salt gradients, as we demonstrate in the following Section.

\subsection{\label{sec:2dDPresults}Diffusiophoresis in 2d Salt Gradients}
The one-dimensional gradients of Section~\ref{sec:1dDPresults} are combined by superposing two different gradients in orthogonal directions. This is achieved using branched microfluidic devices containing two hydrogel sources, as described in Appendix~\ref{sec:microfluidics} and shown in \Fig{fig:Schematic}~(d,~e). The only experimental observable is the particle velocity field, which \Refcite{Warren2020} predicts to arise as a vector sum of chemiphoresis, local electrophoresis, and nonlocal electrophoresis. We measure this velocity field, and compare it to the fields predicted by modelling with and without the current-induced contribution to DP predicted to exist in two dimensions. Three salt combinations are chosen based on their qualitatively different one-dimensional DP behaviour: KOAc ($\beta > 0$) superposed with NaCl ($\beta < 0$); TBAB ($\beta < 0$) superposed with NaCl ($\beta < 0$); and KCl ($\beta \approx 0$) superposed with KOAc ($\beta > 0$). A fourth combination, KCl ($\beta \approx 0$) superposed with NaCl ($\beta < 0$), behaves very similarly to TBAB superposed with NaCl, and is therefore omitted from the article, but included in the Supplementary Information \cite{supplement}.

\subsubsection{Orthogonal Gradients of KOAc and NaCl}
Figure~\ref{fig:2dArrows} (a) shows a snapshot of the experimental velocity field measured at $t = 115\,\mathrm{s}$ after initiation in an experiment superposing a vertical gradient of KOAc with a horizontal gradient of NaCl. The KOAc source is located approximately \SI{840}{\micro\metre} from the top edge of the field of view and the NaCl source is located approximately \SI{865}{\micro\metre} from the right edge of the field of view. \change{A Cartesian co-ordinate system, $(r_\mathrm{NaCl},r_\mathrm{KOAc})$, is defined as illustrated in Fig. \ref{fig:Schematic} (e) and the top right of Fig. \ref{fig:2dArrows}. The co-ordinate $r_\mathrm{NaCl}$ is the normal distance from a line drawn at the edge of the NaCl source and, similarly, $r_\mathrm{KOAc}$ is the normal distance from a line drawn at the edge of the KOAc source. Unit vectors in the $r_\mathrm{NaCl}$ and $r_\mathrm{KOAc}$ directions are perpendicular to one another, and the origin of the co-ordinate system is located between the two sources in experiment, or in the top-right corner in the model.} The location of the microscope field of view relative to the sources varies between experiments, and so \change{the origin of the co-ordinate system is} measured for each experiment using composite images as described in the Supplementary Information \cite{supplement}.

Figure~\ref{fig:2dArrows} (a) reveals that the direction and magnitude of particle velocity are dependent on position relative to the salt sources. At this instant, particles on the right  of the field of view are moving towards the NaCl source with speeds between 1 and \SI{2}{\micro\metre\per\second}. This motion is reminiscent of that expected in a one-dimensional gradient of NaCl, which drives up-gradient DP of these particles. Ahead of this NaCl-attractive front, there is a region in which velocities are smaller and directed upwards, towards the KOAc source. Further still from the NaCl source, velocities are directed towards the upper left corner.

Figure~\ref{fig:2dArrows} (a) shows only the snapshot at $t=115\,\mathrm{s}$, but the velocity field evolves with time, see Movies~1 and~2. As time proceeds, the NaCl-attractive front moves outwards and the velocity vectors rotate past the vertical, until the attraction towards the NaCl source dominates the field of view and all vectors point towards the right. Along with this change in direction, the overall DP speed decreases with time, consistent with the measurements of one-dimensional DP shown in \Fig{fig:1dDP}.

The experimental results, \Fig{fig:2dArrows}~(a), are now compared with the predictions of the full model, \Fig{fig:2dArrows}~(b), and the model without the nonlocal term, \Fig{fig:2dArrows}~(c). Model vector fields are shown over the full simulated region, with the red boxes indicating the the approximate location of the experimental field of view.

Within the red boxes and close to the NaCl source, DP in both full and local-only models is dominated by an up-gradient NaCl-attraction, consistent with the experimental velocity field. However, \change{in the bottom left of the red box in \Fig{fig:2dArrows}~(c), the local-only model predicts that particles are almost stationary. This contrasts with the full model, \Fig{fig:2dArrows}~(b) which qualitatively agrees with the experiment and predicts motion up and to the left. This particle motion can only be reproduced by a model incorporating the nonlocal, current-induced, contribution to DP.} To the best of our knowledge, these data are the first experimental observation of this effect.

\change{The full dynamics may be compared by comparing the vector fields for experimental particle motion in Movie 2 with those predicted by the full (Movie 3) and local-only (Movie 4) models. Only the full model shows the same transient motion up and to the left as observed in experiment.}

\change{However agreement between this experiment and the full model is not quantitative. At its fastest, the motion in this particular experiment is $\sim \SI{4}{\micro\metre\per\second}$ whereas the model predictions are $\sim \SI{1}{\micro\metre\per\second}$. The quantitative discrepancy between experiments and model results likely results from the combined effect of many differences. The geometry of the model does not attempt to accurately replicate the geometry of the branched microfluidic devices  and salt sources. The initial conditions in modeling and experiment are necessarily different. In the experiments an initial liquid flow is needed for the colloidal suspension to flow into the device. This flow will transport some salt away from the sources. In the model, the chemical potential is fixed at the boundaries and so the models tend towards steady-state crossed gradients, while the experiments approach an equilibrium of uniform concentrations. Modeling neglects any advective particle transport in fluid flows. In experiment such flows could occur, driven by diffusioosmosis at the device walls, convection, or any other imperfections in the construction of the microfluidic devices. Despite all these sources of quantitative discrepancy, the qualitative agreement between the experimental observations and the DP model incorporating the nonlocal, current-driven contribution is striking.} 

\begin{figure*}
\includegraphics[width=175mm]{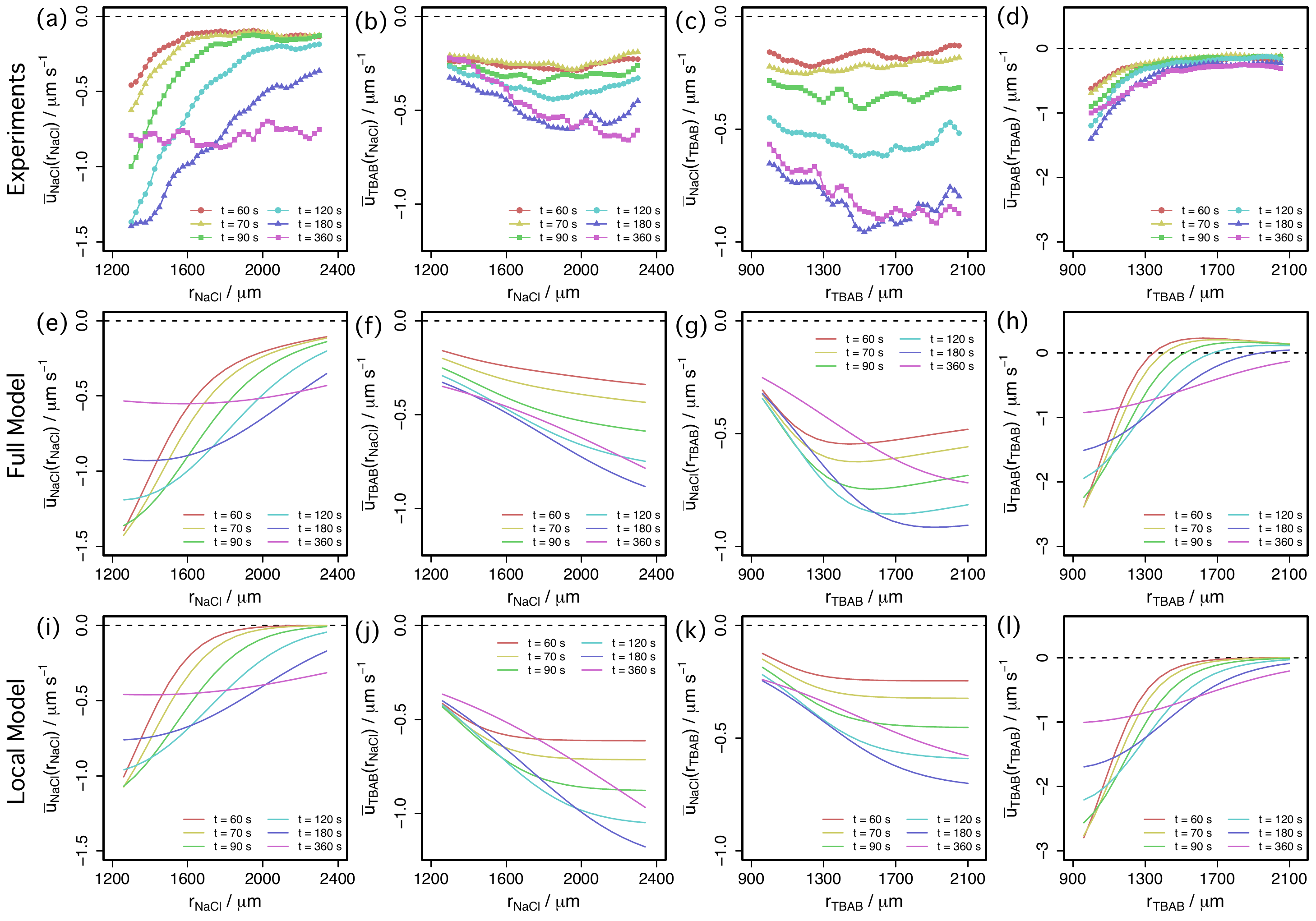}
\caption{\label{fig:TBABNaClVProfs} TBAB superposed with NaCl. Projected velocity profiles in the region for which all three experiments overlap. Colour represents time after initiation, as indicated in legends. Horizontal dashed lines indicate $U=0$. (a-d) Experimental velocity profiles computed as an average over three repeated experiments. (e-h) Velocity profiles predicted by the full model, including the nonlocal contribution to DP. (i-l) Velocity profiles predicted by the local model:
  (a), (e), and (i) show $\bar{U}_\mathrm{NaCl}(r_\mathrm{NaCl})$\,;
  (b), (f), and (j) show $\bar{U}_\mathrm{TBAB}(r_\mathrm{NaCl})$\,;
  (c), (g), and (k) show $\bar{U}_\mathrm{NaCl}(r_\mathrm{TBAB})$\,;
  (d), (h), and (l) show $\bar{U}_\mathrm{TBAB}(r_\mathrm{TBAB})$.}
\end{figure*}

\begin{figure*}
\includegraphics[width=175mm]{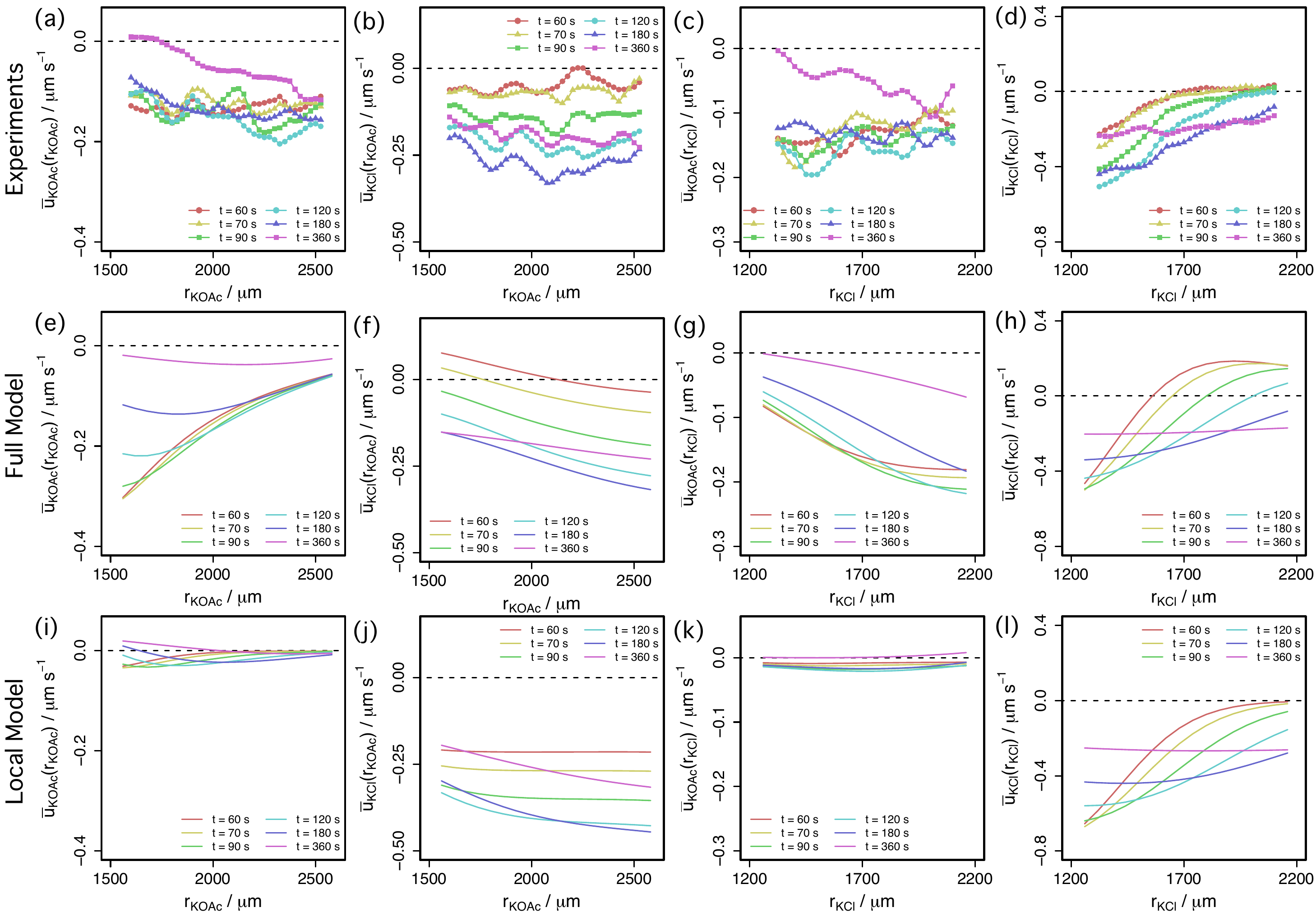}
\caption{\label{fig:KClKOAcVProfs} KCl superposed with KOAc. Projected velocity profiles in the region for which all three experiments overlap. Colour represents time after initiation, as indicated in legends. Horizontal dashed lines indicate $U=0$. (a-d) Experimental velocity profiles computed as an average over three repeated experiments. (e-h) Velocity profiles predicted by the full model, including the nonlocal contribution to DP. (i-l) Velocity profiles predicted by the local model:
  (a), (e), and (i) show $\bar{U}_\mathrm{KOAc}(r_\mathrm{KOAc})$\,;
  (b), (f), and (j) show $\bar{U}_\mathrm{KCl}(r_\mathrm{KOAc})$\,;
  (c), (g), and (k) show $\bar{U}_\mathrm{KOAc}(r_\mathrm{KCl})$\,;
  (d), (h), and (l) show $\bar{U}_\mathrm{KCl}(r_\mathrm{KCl})$.}
\end{figure*}

Comparison of two-dimensional vector fields is challenging, so in \Fig{fig:KOAcNaClVProfs} each velocity field is projected onto four one-dimensional velocity profiles \change{in the $(r_\mathrm{NaCl},r_\mathrm{KOAc})$ co-ordinate system}, which are more easily compared. The three repeated implementations of each experiment are averaged. However, since the location of the field of view differs between experiments, we \change{compute average velocities} only the subregion where all experiments overlap \change{as illustrated in Section X of the Supplementary Information}. Over this subregion, four velocity profiles are computed: \numbr{i} the average velocity towards / away from the NaCl source as a function of distance from the NaCl source, $\bar{U}_\mathrm{NaCl}(r_\mathrm{NaCl})$; \numbr{ii} the average velocity towards / away from the KOAc source as a function of distance from the NaCl source, $\bar{U}_\mathrm{KOAc}(r_\mathrm{NaCl})$; \numbr{iii} the average velocity towards / away from the NaCl source as a function of distance from the KOAc source, $\bar{U}_\mathrm{NaCl}(r_\mathrm{KOAc})$; and \numbr{iv} the average velocity towards / away from the KOAc source as a function of distance from the KOAc source, $\bar{U}_\mathrm{KOAc}(r_\mathrm{KOAc})$. The same quantities are computed from the full and local model velocity fields, in the subregion corresponding to the overlapping region of the three experiments. 

The projected profiles are shown in \Fig{fig:KOAcNaClVProfs} at six times between $t = 60\,\mathrm{s}$ and $t = 360\,\mathrm{s}$. The first row (a--d) shows the average projected velocity profiles computed over three independent experiments. The second (e--h) and third (i--l) rows show the projected profiles predicted by the full and local models, respectively. Negative (positive) velocities represent motion towards (away from) the corresponding electrolyte source. 

From \Fig{fig:KOAcNaClVProfs}~(a), we see that at the earliest times, profiles are positive, indicating an effective NaCl-repulsion. At the latest times, profiles are negative, indicating an NaCl-attraction. The same qualitative behaviour is evident in the model predictions incorporating the current-induced contribution to DP, see \Fig{fig:KOAcNaClVProfs}~(e). By contrast, the local model predicts profiles that are always negative, \ie\ motion is always towards the NaCl source. Our findings for the other three sets of projected velocity profiles are similar, the full model agrees qualitatively with experiment while the local model often predicts velocities with the wrong sign, especially at early times. In the case of motion towards / away from the KOAc source, the local model predicts almost zero velocity due to KOAc's DP-neutrality with our particles, while experiments show clear negative velocities. These negative velocities emerge in the modelling when the current-induced term is included.

In all projections, the differences between the full model (middle row) and local model (bottom row) predictions are most pronounced at earlier times. The current-driven component of DP scales with the reciprocal of conductivity, which is largest when ion concentration is small. In these experiments, for which the concentration gradient is transient, ion concentration is lowest at initiation and increases as ions are released from the hydrogel sources. Therefore, nonlocal DP effects are increasingly suppressed as time progresses, gradients become more shallow and conductivity increases.

\subsubsection{Orthogonal gradients of TBAB and NaCl}
The second combination considered is TBAB superposed with NaCl. Both of these salts have $\beta < 0$, and so both drive up-gradient DP of negatively charged colloids in one-dimensional gradients (\Fig{fig:1dDP}). Movie~5 shows the colloidal motion observed in orthogonal TBAB and NaCl gradients. The projected average velocity profiles measured in experiment and generated by modelling are shown in \Fig{fig:TBABNaClVProfs}. Compared to the previous case of KOAc superposed with NaCl, the differences between the full (middle row) and local (bottom row) model predictions are much less apparent. The model predictions for $\bar{U}_\mathrm{NaCl}(r_\mathrm{NaCl})$ (first column), $\bar{U}_\mathrm{TBAB}(r_\mathrm{NaCl})$ (second column), and $\bar{U}_\mathrm{NaCl}(r_\mathrm{TBAB})$ (third column) show the same qualitative behaviour as one another, and either model could be said to be in agreement with the experimental data (top row). 

The only qualitative difference between the full and local models is in the $\bar{U}_\mathrm{TBAB}(r_\mathrm{TBAB})$ profiles. The local model predicts that $\bar{U}_\mathrm{TBAB}(r_\mathrm{TBAB}) < 0$ for all times, representing a TBAB-attraction. The effect of adding the current-induced contribution to DP is that $\bar{U}_\mathrm{TBAB}(r_\mathrm{TBAB})$ crosses the $U=0$ line and shows a weakly TBAB-repulsive region at early times. The magnitude of this predicted down-gradient DP motion is significantly smaller than the up-gradient, TBAB-attraction predicted closer to the TBAB source. The experimentally measured $\bar{U}_\mathrm{TBAB}(r_\mathrm{TBAB})$ profiles shown in \Fig{fig:TBABNaClVProfs}~(d) do not cross the $U=0$ line and do not show a region of down-gradient, TBAB-repulsive motion.
For this pair of salts, the predicted signature of nonlocal DP is very weak. It is therefore not surprising that the experiments do not reveal it.

\subsubsection{Orthogonal gradients of KCl and KOAc}
The final combination of salts considered is KCl ($\beta \approx 0$) crossed with KOAc ($\beta > 0$). Movie~6 shows the colloidal motion observed in orthogonal KCl and KOAc gradients. The projected average velocity profiles measured in experiment and generated by modelling are shown in \Fig{fig:KClKOAcVProfs}. These are similar to those shown in \Fig{fig:KOAcNaClVProfs} for KOAc superposed with NaCl. This is unsurprising, given that, like NaCl, a one-dimensional KCl gradient drives up-gradient DP of negatively charged colloids (\Fig{fig:1dDP}).

Figure~\ref{fig:KClKOAcVProfs}~(i,~k) show that, without the current-induced contribution to DP, only very slow motion towards or away from the KOAc source is predicted. When the nonlocal term is included, (e) and (g), KOAc-attractive motion is predicted, and this is precisely what is observed in the experimental data, (a) and (c). Furthermore, the full model predicts a transient KCl-repulsive region characterised by $\bar{U}_\mathrm{KCl}(r_\mathrm{KCl})$ crossing the $U=0$ line at early times, \Fig{fig:KClKOAcVProfs}~(h). This signature is observed in the experimental data shown in (d), albeit the measured KCl-repulsive motion is very slow. As was the case for KOAc superposed with NaCl, the experimental observation of these signatures which are unique to the full model is taken as confirmation of the existence of the nonlocal, current-induced contribution to DP predicted by \Refcite{Warren2020}.

\section{\label{sec:conclusion}Conclusion}
We have directly observed the motion of charged colloidal particles in orthogonal concentration gradients of a range of salt pairs. We have compared the results with the predictions of the model described in \Refcite{Warren2020} and found that our observations can only be explained with a nonlocal contribution to DP that is absent in one-dimensional gradients or with only one salt. These are the first experimental observations of this phenomenon and strongly support the theory in \Refcite{Warren2020}. We also experimentally measured DP in one-dimensional concentration gradients of our four salts and verified its consistency with the DPA model. 

The nonlocal effect was strongest with NaCl and KOAc. This pair of salts have very different $\beta$ values ($-0.21$ and $+0.28$), and KOAc has a one-dimensional DP mobility $\DDP$ that is an order of magnitude smaller that of NaCl. Two very different $\beta$ values are essential to generate nonlocal DP; the effect disappears if the two $\beta$ values are equal.

The nonlocal effect is most easily identified at early times, and far from the sources. See Supplementary Movie~2 where the motion of the particles due to the nonlocal term is easiest to see after around 10\% (\ie\ two minutes) of the twenty-minute experimental run shown in the movie, and in the bottom-left corner, farthest away from the two sources of salt. This is also shown in \Fig{fig:2dArrows}. There are two effects at play here that aided us in clearly identifying the nonlocal contribution to DP: first the time scale for the propagation of local DP is much larger than that for nonlocal DP, and second the small value of $\DDP$ of one salt (KOAc).

Starting with the first, the local electrophoresis and chemiphoresis terms are both proportional to \emph{local} gradients in the logarithm of salt concentration. For our millimetre field of view it takes of order $(\SI{1}{\milli\metre})^2/\Da\sim\SI{20}{\minute}$ for salt, and hence local DP, to move across the system, and this can be seen as a front moving from right to left in Supplementary Movie~2, which follows the system for 20 minutes. By contrast, the time scales for the appearance of a current and its associated electric field are all much less than a second (\Fig{fig:timescales}). This is much too fast for us to observe so we detect the nonlocal effect as instantaneous action at a distance on the colloids. For a further discussion, see Appendix~\ref{app:timescales}.

For the second, unambiguously attributing particle motion to the nonlocal effect is easiest when one of the salts (KOAc) has $\beta$ such that the two terms in conventional salt diffusiophoresis are oppositely directed and almost cancel. For example, KOAc provides the vertical gradient in \Fig{fig:2dArrows} and while there is strong motion to the right at the right-hand edge (towards the NaCl source) there is no vertical motion along the top edge, towards or away from the KOAc source. This cancellation of chemiphoresis and local electrophoresis allows the nonlocal term to dominate, making its effect easier to identify. 

However, this cancellation does not need to be perfect. Additional modelling (see Supplementary Information \cite{supplement}) of the salt pair sodium tetraphenylborate (NaTPB, $\beta = 0.4$ \cite{Williams2022}) and NaCl also predicts a clearly measurable nonlocal effect. NaTPB has larger $\beta$ than KOAc, meaning that cancellation of the two local terms is much less good than for KOAc. We cannot verify this prediction in experiment as NaTPB inhibits the ability of the PEG-DA precursor solution to form a hydrogel.

\change{Another advantage of nonlocal DP is due to the fact that it can occur in locations where there is no salt gradient. The local driving force for convection is the mass density gradient which will be absent at points where there is no concentration gradient. So nonlocal DP can occur in locations in a solution where the local driving force for convection is zero. In our system we suppressed convection --- to make it easier to see DP --- by making the vertical dimension small. However, it may be possible to see nonlocal DP in systems where salt gradients do drive significant convection but this convection is far from where there is nonlocal DP.}

There are several differences between the experiments and the model. The model geometry and boundary conditions do not perfectly recreate those of the experiments. The electrolyte sources in the model span the entire top and right edges of the square domain, while in experiment, the sources have a finite width. Furthermore, the model boundaries are maintained at constant chemical potentials, meaning that the models are approaching a steady state concentration gradient. By contrast, the experiments are evolving towards an equilibrium characterised by uniform concentration.  Nevertheless the concordance between experiments, and predictions combining the electroneutral Nernst-Planck equations and the DPA theory of DP, is naturally very pleasing. 

\change{Additional phenomena may contribute to the motion of the colloids in experiments which are excluded by design in modelling, where colloidal motion is determined entirely by diffusiophoresis. Both natural convection due to the dependence of mass density on solution concentration and diffusioosmosis at the top and bottom walls of the  device can generate fluid flows that move particles. In one-dimensional concentration gradients in sealed devices, both of these generate circulating flows of known form \cite{Williams2020,Gu2018}. Such flows would add a height-dependent advective component to colloidal motion, in addition to the height-independent diffusiophoresis.
Advection of particles in these circulating flows due to diffusioosmosis or natural convection is expected to drive particles in different (opposite) directions at different heights, which we do not see significant evidence of. 
Hence it is likely that diffusioosmosis is weaker than the motion due to diffusiophoresis. But the flows should be present, and, we assume, account for some of the quantitative discrepancy between modeling and experiment. Diffusiophoresis and diffusioosmosis have the same source so it is typically not possible to eliminate one entirely while keeping the other. We estimate the maximum convection flow speeds in Section VII of the Supplementary Information and find that they should be less than $\SI{0.1}{\micro\metre\per\second}$ which is at the limit of what we can detect in experiment.
}

Because of its effectively instantaneous propagation speed, nonlocal DP opens new applications where local DP would take too long to start up. The DP velocity also does not have to be towards, or away from, salt sources. This in contrast to a concentration gradient of a single salt, where DP is necessarily directed either up or down the local gradient \cite{Velegol2016, Paustian2015, Shin2016, Banerjee2016, Banerjee2019}.  We hope that this work inspires future research into using DP to move particles with faster start ups, and in more complex, more controlled ways.

\section*{Data Availability}
Experimental and modelling data supporting this article are freely available at Figshare \cite{FigshareData} \url{https://doi.org/10.6084/m9.figshare.23579262.v1}.

\begin{acknowledgments}
The authors thank Anirudha Banerjee for advice regarding soluto-inertial beacons and Bertie Woodward-Rowe for advice regarding hydrogel formation and measurements of the UV lamp spectrum and power density. Funding for IW was provided by the EPSRC through a New Horizons grant (Grant No. EP/V048473/1). 
\end{acknowledgments}









\appendix

\section{\label{sec:materials}Materials}
Sodium chloride (NaCl, 99.9\% pure), potassium acetate (KOAc, 99.0\% pure) and tetrabutylammonium bromide (TBAB, 98.0\% pure) were purchased from Sigma Aldrich and used as received. Potassium chloride (KCl, 99\% pure) was purchased from Fisher Scientific and used as received. Fluorescently labeled polystyrene particles of diameter \SI{1}{\micro\metre} and $\zeta$-potential $\approx -50\,\mathrm{mV}$ \cite{Yaehne2013} (Invitrogen FluoSpheres\texttrademark{}, carboxylate-modified, yellow-green fluorescent) were purchased from ThermoFisher as a 2\% solids suspension and diluted to 1 drop ($\approx\SI{30}{\micro\liter}$) per 5\,mL. Poly(ethylene glycol) diacrylate (PEG-DA), number-average molecular weight $M_n = \SI{700}{\gram\per\mol}$ was obtained from Merck. Photoinitiator 2-hydroxy-2-methylpropiophenone was obtained from Sigma Aldrich.

\section{\label{sec:microfluidics}Microfluidic Device Fabrication}
Single-use microfluidic devices are made by sandwiching double-sided adhesive tape (VK3220, Viking Industrial) between a microscope slide and a coverslip \cite{Nath2010, Khashayar2017} as shown in \Fig{fig:Schematic}~(a). \change{Good adhesion between the tape and the glass is ensured by clamping the device with a pair of bulldog clips and placing it in an oven at $80^\circ \, \mathrm{C}$ for 20 minutes.} Channel depth is set by the tape thickness and is approximately \SI{50}{\micro\meter}, which is sufficiently shallow to suppress circulating convective flows due to the mass density gradient associated with a concentration gradient \cite{Williams2020}. \change{This is justified in more detail in Section VII of the Supplementary Information.} Channels are either T-shaped, \Fig{fig:Schematic}~(b,~c), for studies of one-dimensional DP with a single salt source, or have a branched shape, \Fig{fig:Schematic}~(d,~e), for studies of two-dimensional DP in crossed gradients with sources of two salts. 

Ionic gradients are created by adapting the soluto-inertial beacons of Banerjee \etal\ \cite{Banerjee2016, Banerjee2020}. This approach uses fixed hydrogel structures containing a high concentration of solute. When a low concentration solution is brought into contact with the hydrogel, solute effluxes at a rate set by the diffusive and associative properties of the hydrogel and solute \cite{Banerjee2019PRE}. 

A hydrogel precursor stock solution is created by mixing deionised (DI) water with 40\% by volume PEG-DA and 4\% by volume photoinitiator. This stock is mixed at a 1:1 volume ratio with a $260\,\mathrm{mM}$ salt solution to create a precursor solution containing 20\% by volume PEG-DA, 2\% by volume photoinitiator, and $130\,\mathrm{mM}$ salt. The source channel(s) of the microfluidic devices are filled with this precursor \change{via capillary action, carefully adding one drop at a time to avoid overfilling the source channel and leaking out past the channel intersection.} \change{The device is then} illuminated with UV light for 60 seconds to crosslink the PEG-DA and form a hydrogel. The UV source spectrum has power density of approximately $5\,\mathrm{mW}\,\mathrm{cm}^{-2}$ at $365\,\mathrm{nm}$.

\section{\label{sec:microscopy}Microscopy and Image Analysis}
The suspension of fluorescent polystyrene particles is loaded by capillary action into the device through one end of the sample channel (one-dimensional devices) or through the inlet channel between the two sources (two-dimensional devices). \change{As the suspension is drawn into the device from the inlet, air is forced out via the outlet(s). It is important to expel all the air and not trap any bubbles in the device.} The open \change{external} ends of \change{inlet, outlet, and source channels} are sealed using Araldite Rapid 5 Minute epoxy adhesive \change{at the locations indicated by the green, dashed ovals in Fig. \ref{fig:Schematic} (b) and (d). The device is therefore isolated during data acquisition rather than connected to an external reservoir of the particle suspension.} 

The sample is placed on an upright microscope (Olympus BX3M) operating in epifluorescence mode with a $10\times$ magnification objective. In one-dimensional experiments, the hydrogel source is included in the microscope field of view, while in two-dimensional experiments, the hydrogel sources are located outside of the field of view, as indicated by the green region in \Fig{fig:Schematic}~(e). The field of view spans a region of approximately \SI{1700}{\micro\metre} by \SI{1300}{\micro\metre}. Images of resolution 1224 by 960 pixels are acquired at 1 frame per second for 20 minutes. \change{The time between loading the device and beginning acquisition is approximately 60 seconds. Within this time, residual flows resulting from the loading procedure have subsided to speeds of $\lesssim 0.1 \, \mu \mathrm{m} \mathrm{s}^{-1}$, as demonstrated by an experiment performed with two hydrogels formed without any salts, described in the Supplementary Information.}

During an experiment, a subpopulation of the particles adhere to the lower glass surface and appear stationary. Including stationary particles in subsequent data analysis would lead to an underestimate of DP speed and they are therefore removed by subtracting the time-averaged image over the whole acquisition period from each frame of the video. However, in two-dimensional experiments, the stationary particles are useful for aligning composite images used to locate the microscope field of view relative to the hydrogel sources, as described in the Supplementary Information \cite{supplement}.

The particle velocity field is quantified using particle imaging velocimetry (PIV) \cite{Westerweel1997, Adrian2005} implemented in ImageJ \cite{Tseng2012}. This approach estimates a velocity field through two-dimensional correlations in image intensity in two images separated by a time interval $\Delta t$. An image at time $t_0$ is divided into interrogation regions, and each interrogation region is compared to the image at later time $t_0 + \Delta t$. The displacement of the interrogation region which maximises the correlation between the two images gives an estimate of the local velocity. Repeating this for each interrogation region and each pair of images gives a velocity field that evolves in time. Here, overlapping square interrogation regions of side length 64 pixels ($\approx\SI{90}{\micro\metre}$) centred on a square grid of locations separated by 32 pixels ($\approx\SI{45}{\micro\metre}$), and image pairs separated by $\Delta t = 5\,\mathrm{s}$ are employed. 

In the absence of DP, particles exhibit Brownian motion, and under this scenario it is incorrect to assume that the PIV velocity in one interrogation region is correlated to the velocity in an adjacent region. Consequently, to ensure the procedure is agnostic to the existence of DP, no normalised median or dynamic mean test is performed in post-processing of the PIV velocity fields. Instead, noise is suppressed by applying a moving average over a time interval of 10 seconds. 

\section{One-dimensional Diffusion Model}\label{sec:app1d_diff}

\begin{figure}[htb]
\includegraphics[width=65mm]{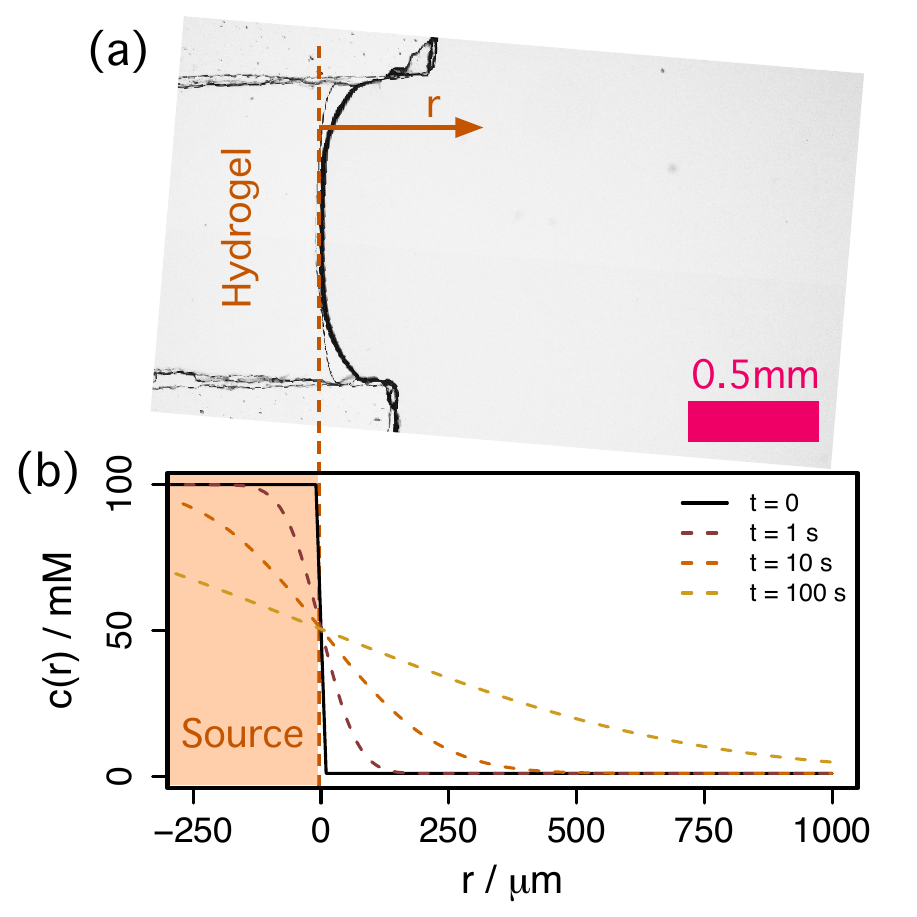}
\caption{\label{fig:1DModel} \change{One-dimensional diffusion model. The experimental geometry shown in the micrograph in (a) is reduced to a 1-dimensional diffusion problem in the co-ordinate $r$ representing the normal distance from the edge of the hydrogel. The micrograph is rotated to align $r$ horizontally. (b) One dimensional concentration profiles according to Eq. \ref{eq:conc1d} computed for NaCl. The shaded region represents the inside of the hydrogel source. The edge of the source is located at $r=0$. The solid black curve represents the initial condition and dashed curves illustrate the time evolution of concentration, according to the legend.}}
\end{figure}

\change{The modeling of our 1-dimensional DP experiments is illustrated in Fig. \ref{fig:1DModel}.} The diffusion equation in one dimension for salt concentration is
\begin{equation}
\label{eq:1dDiffusion}
\frac{\partial c(r,t)}{\partial t} = \Da\, \frac{\partial^2 c(r,t)}{\partial r^2}\,.
\end{equation}
The salt is treated as a single species which diffuses at a rate given by the ambipolar diffusion coefficient $\Da = 2D_+ D_-/(D_+ + D_-)$, which accounts for the coupling between anion and cation diffusion \cite{Gupta2020}. The initial condition is modeled as a step function, \change{shown as the solid black line in Fig. \ref{fig:1DModel} (b),}
\begin{equation}
  c(r,0)=\begin{cases}
  c_s\,,   &  (r < 0) \\
  (c_s + c_0)/2\,, &  (r = 0) \\
  c_0\,,   &  (r > 0)   
  \end{cases}
\end{equation}
with $c_s$ the salt concentration in the hydrogel source, $c_0$ the initial concentration outside of the source, and $c_s \gg c_0$. The boundary conditions are $c(-\infty,t)=c_s$ and $c(\infty,t)=c_0$. \change{This models an idealised, strictly 1-dimensional version of experiments such as are shown in Fig. \ref{fig:Schematic} (c) and Fig. \ref{fig:1DModel} (a), in which salt concentration depends only on time and the normal distance from the edge of the hydrogel, $r$.}

The diffusion equation is then solved by
\begin{equation}
\label{eq:conc1d}
c(r,t) = \frac{c_0 + c_s}{2} + \frac{c_0 - c_s}{2} \, \mathrm{erf} \left( \frac{r}{\sqrt{4\Da t}} \right),
\end{equation}
with spatial derivative
\begin{equation}
\label{eq:dcdr1d}
\frac{\partial c(r,t)}{\partial r} = \frac{c_0 - c_s}{\sqrt{4 \pi \Da t}} \, \exp \left( -\frac{r^2}{4\Da t} \right).
\end{equation}

The concentration immediately outside the source--sample interface, which is the effective value of $(c_s + c_0)/2$, depends on the association constant characterising the affinity between the ions and the hydrogel, and the ion diffusion coefficients within the hydrogel \cite{Banerjee2019PRE}. These quantities are unknown.  Diffusiophoresis depends on the gradient of the logarithm of the concentration so is determined by ratios.  For present purposes we assume the ratio between the source and background concentrations is 100:1. Therefore, we can set the concentration far from source at $c_0 = 1\,\mathrm{mM}$ to fix the units and model one-dimensional gradients of the form in \Eq{eq:conc1d} using an estimated source concentration $c_s = 100\,\mathrm{mM}$. \change{The time evolution of Eq. \ref{eq:conc1d} is shown for NaCl in Fig. \ref{fig:1DModel} (b).} To translate concentration gradients to velocities the values of $\DDP$ in Table~\ref{tab:DDP} are used. Thus, by choosing $c_s$ and $c_0$, $U(r,t) = \DDP \nabla \ln c(r,t)$ is estimated for each salt.

\section{\label{sec:modelmethods}Computational Methods}
\begin{figure}[htb]
\includegraphics[width=70mm]{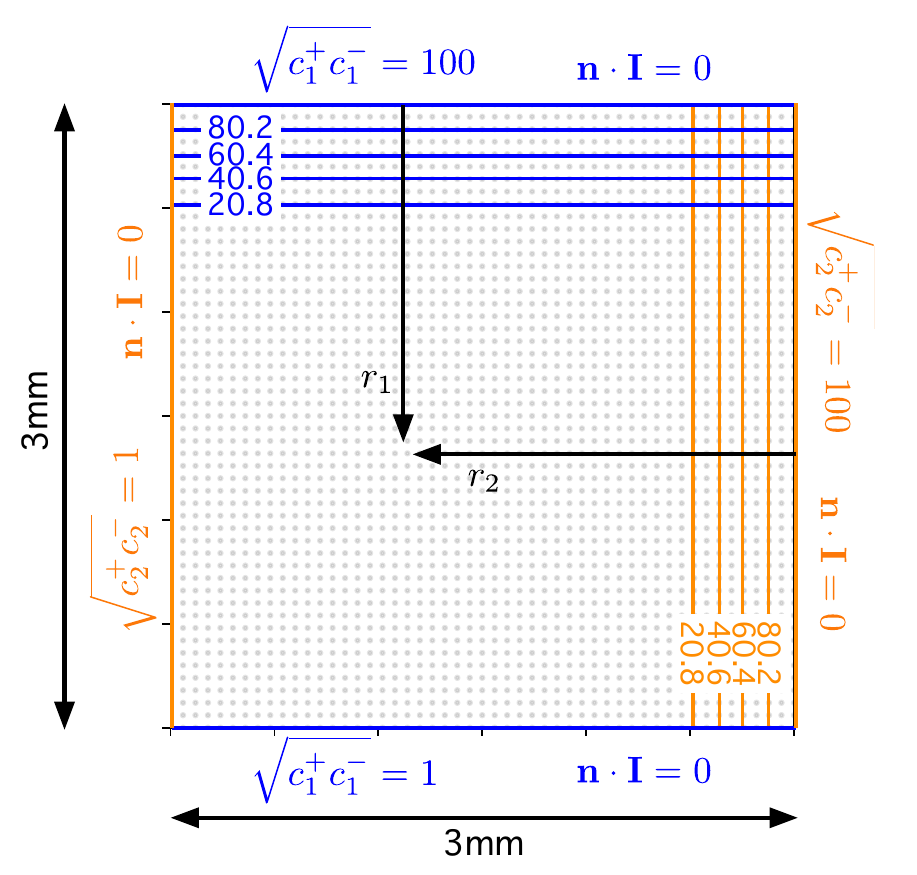}
\caption{\label{fig:2DModel} \change{Schematic showing the initial and the boundary conditions of the two-dimensional model. Ion concentrations are defined on a $50 \times 50$ grid represented by the grey dots. Blue (orange) contours show the initial concentration gradient of salt 1 (salt 2). Constant chemical potential boundary conditions fix the product of salt 1 ion concentrations at the top and bottom edges and the product of salt 2 ion concentrations at the left and right edges. The boundary condition on the current at all edges is $\mathbf{n} \cdot \mathbf{I}=0$ where $\mathbf{n}$ is the unit normal to the edge. Cartesian co-ordinates $(r_1,r_2)$ are defined from the top and right edges.}}
\end{figure}

Two-dimensional DP is modelled computationally using methods developed for \Refcite{Warren2020} \change{and code available at \Refcite{WarrenGithub}}. \change{The simulated geometry, and initial and boundary conditions are illustrated in Fig. \ref{fig:2DModel}.} A square domain of side length $L = \SI{3000}{\micro\metre}$ is considered. The model is initiated with crossed error function concentration distributions of two salts in the $x$ and $y$ directions. The error function in the $x$-direction ($y$-direction) is centred \SI{300}{\micro\metre} from the right (top) edge of the domain and is \SI{300}{\micro\metre} wide. \change{These are represented by the blue and orange contours in Fig. \ref{fig:2DModel}.} The initial concentration at the right (top) edge is 100 times that at the left (bottom) edge, and these edge concentrations are maintained by fixed chemical potential boundary conditions. \change{As described in Supplementary Material to \Refcite{Warren2020}, fixing the chemical potential at the boundary amounts to fixing the product of the anion and cation concentrations for a neutral ion pair. The boundary condition on the current is that the current normal to each edge vanishes.}

The model is iterated through time in $90\,\mathrm{ms}$ steps by numerically solving the Nernst-Planck equations on a $50 \times 50$ grid subject to the condition of solenoidal current, $\nabla \cdot \Ivec = 0$. The model is iterated through $4000$ steps, corresponding to $6$ minutes. 

\change{To facilitate comparison with experiment and to account for the finite width of the initial concentration conditions, the first output of the model is considered to represent $t = 15 \, \mathrm{s}$. This offset time is the time required for a typical ion with $D = 1.5 \times 10 ^{-9} \, \mathrm{m}^2 \, \mathrm{s}^{-1}$ to establish a $300 \, \mu \mathrm{m}$ wide error function concentration profile from a step function at $t=0$ within the 1d problem described in Appendix \ref{sec:app1d_diff}. Initial conditions of finite width are chosen to ensure that the initial concentration gradients are well-represented over multiple grid points. The impact of the choice of initial condition width is discussed in the Supplementary Information.} 

Using the two-dimensional concentration fields of each ion, the three contributions to the DP velocity of a particle with $\zeta = -50\,\mathrm{mV}$ are computed according to \Eq{eq:2ddp}. Depending on the choice of crossed electrolytes, there are either 3 or 4 unique ion species. A complete description of the computational methods is provided in \Refcite{Warren2020}; see also \Refcite{WarrenGithub}

\section{\label{app:timescales}Time scales}
A number of time scales, spanning a large dynamic range, are relevant to DP. These are outlined here, in approximate order from fastest to slowest.  In the calculations we assume an ambipolar diffusion coefficient for ions $\Da\sim10^{-9}\,\msqpersec$ (Table~\ref{tab:Ds}),  the system size $L\approx10^{-3}\,\mathrm{m}$ and solvent viscosity $\nu\approx10^{-6}\,\msqpersec$.
We use $\csalt$ to denote a generic salt (ion) concentration and take the Debye length $\lD\sim10^{-9}\,\mathrm{m}$ to correspond to $\csalt\approx100\,\mathrm{mM}$.

\paragraph{Time scale for electromagnetic waves to propagate across the system} Light, travelling at a speed of order $10^8\,\mathrm{m}\,\mathrm{s}^{-1}$, takes of order $\SI{10}{\pico\second}$ to cross the system. This is extremely short meaning that one can assume electrostatics holds in these problems on longer time scales.

\paragraph{Time scales for ions and colloids to respond to an electric field}  When the local electric field changes, it takes time for the ions to respond and reach a new steady velocity. For an ion of radius $\rion<1\,\mathrm{nm}$ this time scale is order $\rion^2/\nu\alt\SI{1}{\pico\second}$.  This is also very fast, and on longer time scales means that the steady-state assumption in the Nernst-Planck equations can be considered to be valid.  On the other hand the corresponding time scale for micron-sized colloids  (the momentum relaxation time) is of the order microseconds.  This is the time it takes for the colloid to reach its steady-state drift velocity in phoresis.  Since this is also fast, it validates the use of the DPA theory on the salt diffusion time scale (below).

\paragraph{Time scale for formation of electric fields}
We next turn to the question of the time scales for the formation of electric fields in the system, due to space charges or EDL charging.  The physics here is discussed in detail by Bazant \etal\ \cite{bazant2004}; see also \Refscite{Warren2020} and \cite{Warren2023}.  We outline the essential scaling arguments.

Our problem is essentially an electrostatic one (see {\it a.}~above), so the electric field obeys $\nabla\cdot\Evec=-\rho_Q/\epsilon$, where $\rho_Q$ the space charge density. The magnitude of $\Evec$ is at most of order the characteristic thermal potential difference $\kT/e$, divided by the system size $L$.  So the left-hand side of this equation scales as $\kT/(eL^2)$. The space charge density associated with this varying electric field is then $\rho_Q\sim \epsilon\kT/(eL^2)\sim\lD^2ce/L^2$, since the Debye length $\lD^2\sim\epsilon\kT/(ce^2)$.  In units of the elementary charge, this is a tiny fraction ($\lD^2/L^2\ll 1$) of the salt ion concentration.  Thus establishing the space charge density requires only a few ions to move a small distance and so is many orders of magnitude faster than the salt diffusion time itself.
Assuming sample volume $hL^2$, where $h$ is the chamber height and $L$ the lateral size, the time scale can be estimated from the ratio of the charge $Q\sim\rho_QhL^2\sim h\lD^2ce$ needed to establish the electric field, to the current $I= V/R$ (not to be confused with the current \emph{density} in the main text), as $\tau\sim Q/I$ (if we write $Q=CV$, this is equivalent to an $RC$-circuit charging time \cite{bazant2004}).  The voltage $V\sim\kT/e$ as before, and the resistance $R\sim L\varrho/(Lh)\sim(h\sigma)^{-1}$ where the conductance $\sigma\sim \mu ce^2$ in terms of the ionic mobility $\mu=\Da/\kT$.  Hence $R\sim\kT/(h\Da ce^2)$ and $I=V/R\sim h\Da ce$.  Finally, $Q/I \sim h\lD^2ce/(h\Da ce) \sim \lD^2/\Da \sim 10^{-9}\,\mathrm{s}$.  This is the origin of the \change{space-charge formation} time scale in \Fig{fig:timescales}.

The other relevant time scale is the time to charge the electric double layers (EDLs).  This is a factor of $L/\lD$ larger than the space charge formation time \cite{bazant2004}.  The reason is that the amount of charge that must be transferred is $L/\lD$ larger.  To see this, consider that the total charge in the EDLs will be of order the charge per unit area $\lD ce$ (the charge density in the EDL times the EDL thickness), multiplied by the surface area $hL$ (\ie\ the perimeter of the sample chamber).  This gives the estimate $Q\sim hL\lD ce$, which is indeed a factor $L/\lD$ larger than above.  Then, the same line of argument can be followed as for the space charge: the potential, resistance, and current estimates remain the same, making the EDL charging time of the order $L/\lD\times \lD^2/\Da \sim L\lD/\Da \sim 10^{-3}\,\mathrm{s}$.  This is the next time scale shown in \Fig{fig:timescales}. 

These time scales are much faster than the salt diffusion time (below).  Thus we can consider that the space charge giving rise to the electric fields, and the EDLs which control the boundary conditions, are fully formed on \change{salt diffusion} time scale.

\paragraph{Time scale for viscous flow relaxation} The suspension of colloidal tracers is loaded into the experimental microfluidic device as described in Appendix~\ref{sec:microscopy}.  The relaxation time for residual flows is set by the time scale for momentum to diffuse across the \change{shortest distance to a wall; walls function as momentum sinks. The shortest distance is the height $h\simeq\SI{50}{\micro\metre}$, so the timescale is $h^2/\nu$, which is of order milliseconds. This is comparable to the EDL charging time, see above and as shown in \Fig{fig:timescales}.  Again this timescale is} fast compared to the experimental time window. Therefore, one can consider that any residual flows from loading the device will have decayed before data acquisition starts. \change{Flowing the colloidal suspension into the microfluidic device to fill it takes a few seconds.}

\paragraph{Time scale for salt diffusion and local DP}
We now consider the longest remaining time scale in the problem, which is the time it takes for the salt gradients themselves to relax.  This time scale is order $L^2/\Da\sim10^3\mathrm{s}\approx16\,\mathrm{minutes}$, as shown in \Fig{fig:timescales}.  In a nutshell, salt takes tens of minutes to diffuse across the field of view of our microscope.

An important consideration though is that this means that local DP also propagates at the same rate.  To see this, note that both local DP terms scale as $\nabla\ln c$, see \Eq{eq:PrieveAnderson}.  Using \Eqs{eq:conc1d} and \eqref{eq:dcdr1d} for the concentration profile of a diffusing salt we have that
\begin{equation}
\label{eq:dlncdr1d}
-\frac{\partial\ln c(r,t)}{\partial r}  = 
\frac{1}{\sqrt{\pi \Da t}}\times
\frac{\exp ( -{r^2}/{4\Da t})}%
     {\gamma-\mathrm{erf} ( {r}/{\sqrt{4\Da t}})}\,,
\end{equation}
where $\gamma=(c_s+c_0)/(c_s-c_0)$.  A crucial point is that $\gamma>1$ as long as some background ions are present ($c_0>0$), and the denominator tends to $\gamma-1>0$ for $r\to\infty$.  In water, a background ion concentration will always be present due to the dissociation of water molecules to form hydrogen and hydroxyl ions. Even pure water at pH 7 has an ionic concentration of $10^{-7}$moles/litre. 

Under these conditions, for distances $r\gg\sqrt{\pi \Da t}$ from the source, \Eq{eq:dlncdr1d} simplifies to
\begin{equation}
\label{eq:dlncdr1d-2}
-\frac{\partial\ln c(r,t)}{\partial r}
\approx \frac{\exp(-{r^2}/{4\Da t})}{(\gamma-1)\sqrt{\pi \Da t}}\,.
\end{equation}
The Gaussian in here means that when $r\gg\sqrt{\pi \Da t}$ the derivative of $\ln c$ is essentially zero. 
Thus both local DP terms are negligible whenever $r\gg\sqrt{\pi \Da t}$ and they only become appreciable when $r\sim\sqrt{\pi \Da t}$. 

The implication is that when a source of salt is introduced into a system, local DP propagates away from the source at a speed set by salt diffusion, and so reaches a distance $r$ after a time $r^2/\Da$. This is true even for the local electric field term, which depends on the local value of $\nabla\ln c$ which propagates via diffusion. This contrasts completely with the nonlocal electric field which comes into existence when the space charge forms, on a much faster time scale.  The significance is that at least at early times when $\sqrt{\Da t}\ll L$, local DP will be confined to regions around the ion sources, and outside these regions DP will be due to the nonlocal action-at-a-distance effect.  The \caveat\ is that this effect requires crossed salt gradients somewhere in the system, so that unless the ion sources are close together, there is only a time window of order $L^2/\Da$ where nonlocal DP can be cleanly observed (\ie\ the ions have to diffuse far enough for the gradients to meet, but not so far that they swamp the whole system).  Fortuitously, this is exactly the time window probed by our experiments.


%

\end{document}